\documentclass[a4paper, 10pt, twoside, english]{article}

\usepackage[utf8]{inputenc}
\usepackage[T1]{fontenc}
\usepackage[sc]{mathpazo}
\usepackage[english]{babel}
\usepackage{geometry}
\usepackage{amsmath,amsfonts,amssymb,amsthm}
\usepackage[final,colorlinks]{hyperref}
\usepackage{graphicx}
\usepackage{cleveref}
\hypersetup{
     colorlinks=true,
     linkcolor=black,
     filecolor=blue,
     urlcolor=Mahogany,
     }
\usepackage{mathtools}
\usepackage[dvipsnames]{xcolor}
\usepackage{pdflscape}
\usepackage{etoolbox}
\usepackage{fancyhdr}
\usepackage{lastpage}
\usepackage{ifdraft}
\usepackage[autostyle,italian=guillemets]{csquotes}
\usepackage[backend=bibtex, maxnames=50, firstinits=false, sorting=nyt]{biblatex}
\usepackage{colortbl}
\usepackage{booktabs}
\usepackage{hyphenat}
\usepackage{siunitx}
\usepackage[mathlines,pagewise]{lineno}

\graphicspath{{./images/}}

\usepackage{colortbl}
\usepackage{amssymb}
\usepackage[shortlabels]{enumitem}
\setlist[enumerate,1]{label=\textnormal{(\emph{\roman*})}}
\usepackage{cases}

\definecolor{amber}{rgb}{1.0, 0.75, 0.0}
\definecolor{aogreen}{rgb}{0.0, 0.5, 0.0}
\definecolor{darksienna}{rgb}{0.24, 0.08, 0.08}

\allowdisplaybreaks

\usepackage{array,multirow}

\usepackage{changes} 
\setdeletedmarkup{\color{red}\sout{#1}}

\newcommand{\dd}{\mathop{}\!\mathrm{d}}
\DeclareMathOperator*{\e}{e}

\newcommand{\R}{\mathbb{R}}
\newcommand{\N}{\mathbb{N}}

\theoremstyle{plain}

\theoremstyle{definition}

\numberwithin{equation}{section}
\numberwithin{table}{section}
\numberwithin{figure}{section}

\usepackage[shortlabels]{enumitem}
\setlist[enumerate,1]{label=\textnormal{(\emph{\roman*})}}

\fancypagestyle{mypagestyle}{%
  \fancyhf{}%
  \fancyhead[LO,RE]{\scriptsize Approximation of reproduction numbers for age-structured models}%
  \fancyhead[RO,LE]{\scriptsize S. De Reggi, F. Scarabel, R. Vermiglio}%
  \fancyfoot[LE,RO]{\scriptsize \thepage\,/\,\pageref*{LastPage}}%
}
\usepackage{authblk}
\usepackage{titling}

\usepackage{footnote}
\thanksmarkseries{arabic}

\title{Approximating reproduction numbers: a general numerical method for age-structured models}

\author{Simone De Reggi\thanks{CDLAb - Computational Dynamics Laboratory, University of Udine, Italy}\thanksgap{2mm}$^,$\thanks{Department of Mathematics, Computer Science and Physics, University of Udine, Italy}\thanksgap{2mm}$^,$\thanksmark{4}
,\ Francesca Scarabel\thanksmark{1}\thanksgap{2mm}$^,$\thanks{Department of Applied Mathematics, University of Leeds, United Kingdom}\thanksgap{2mm}$^,$\thanksmark{4},\ Rossana Vermiglio\thanksmark{1}\thanksgap{2mm}$^,$\thanksmark{2}\thanksgap{2mm}$^,$\thanks{e-mail: \url{simone.dereggi@uniud.it},  \url{f.scarabel@leeds.ac.uk}, \url{rossana.vermiglio@uniud.it}}}
\date{March 7, 2023}

\begin{document}
\maketitle
\begin{abstract}
\noindent In this paper, we introduce a general numerical method to approximate the reproduction numbers of a large class of multi-group, age-structured, population models with a finite age span. To provide complete flexibility in the definition of the birth and transition processes, we propose an equivalent formulation for the age-integrated state within the extended space framework. Then, we discretize the birth and transition operators via pseudospectral collocation. We discuss applications to epidemic models with continuous and piecewise continuous rates, with different interpretations of
the age variable (e.g., demographic age, infection age and disease age) and the transmission terms (e.g., horizontal and vertical transmission). The tests illustrate that the method can compute different reproduction numbers, including the basic and type reproduction numbers as special cases.


\end{abstract}

\section{Introduction}

Reproduction numbers are key quantities in epidemiology, as they are usually related to concepts such as the occurrence of epidemic outbreaks, the herd immunity level, the final size, and the endemic equilibrium \cite{Anderson1991}. They were originally introduced in the context of demography and ecology, where they typically characterize either population persistence or extinction \cite{Heesterbeek2002}. 
 The most known and used example of a reproduction number in epidemiology is the \emph{basic reproduction number} (or ratio) $R_0$, which describes the average number of secondary cases produced by a typical infected individual during its entire infectious period, in a completely susceptible population \cite{diekmann1990}. For deterministic models, $R_0=1$ is a threshold that determines the stability of the disease-free equilibrium: it is stable when $R_0<1$ and unstable when $R_0>1$. In simple models, $R_0$ also typically characterizes the herd immunity threshold (i.e., the proportion of population that should be immunized to prevent the spread of the disease) via the expression $1-1/R_0$. Variations of the basic reproduction number when the population is partially immune or when transmission is affected by control measures are also known as \emph{effective} and \emph{control} reproduction numbers \cite{Pellis2022}. 
 
 For more complicated models (e.g., with heterogeneity), looking at $R_0$ alone may not be satisfactory for epidemic control \cite{Heesterbeek2003}. For instance, this is true when it is possible to apply control measures only to a certain group of individuals (e.g., mosquitoes rather than humans in vector-borne infections) or against certain transmission routes (e.g., vertical rather than horizontal transmission). 
 Under these circumstances, a simple and explicit relation between $R_0$ and the herd immunity level may no longer exist \cite{Heesterbeek2003}. In this case, the \emph{type reproduction number}, usually denoted by $T$, is a more appropriate measure, as it describes the expected number of secondary cases in individuals of a certain type produced by one infected individual of the same type, either directly or through chains of infection passing through any sequence of the other types \cite{HEESTERBEEK20073}. As such, $T$ can be directly linked to the amount of control measures to be applied to one specific group of individuals to stop the spread. 
An extension of the type reproduction number, the \emph{state reproduction number}, was proposed in \cite{Inaba2008}, which allows for focus types not only states describing new infections but potentially any epidemic state (e.g., asymptomatic phase and symptomatic phase) \cite[Section 5.2]{Inaba2017}.
Finally, \cite{lewis2019, target2013} considered a further generalization of these quantities, the \emph{target reproduction number}, which focuses on specific interactions between types, rather than all interactions that involve a given type of individuals.
It is important to underline that although different reproduction numbers have different biological interpretations, they typically share the same threshold for epidemic extinction with $R_0$ \cite{lewis2019, Thieme2009}.

\medskip
For deterministic models formulated as ordinary differential equations, a well-established and widely-used framework to compute $R_0$ is that of linearizing the model around the disease-free equilibrium and then computing the spectral radius of a \emph{Next Generation Matrix} (NGM) \cite{diekmann1990, NGM2010}. The method is based on the idea of interpreting infection transmission as a demographic process, where a new infection is considered as a \emph{birth} in the demographic sense.
Intuitively, the NGM maps the distribution of infected individuals in one generation to the distribution of newly infected individuals in the next generation \cite{diekmann1990}; this is mathematically derived from the model's coefficients by suitably splitting the Jacobian into two parts, one accounting for the birth/infection processes and one accounting for the \emph{transition} processes (which include, for instance, changes in the epidemiological state, death, or acquisition of immunity) \cite{driessche2002}. The simplicity of the NGM method has considerably increased its popularity in the last decades. At the same time, the potentially arbitrary splitting into birth and transition processes described above has given rise to many criticisms and misconceptions about $R_0$ and its generational interpretation; see for example \cite{failure}. In fact, the interpretation of birth and transition is typically left to the modeller \cite{didactic} and, while different choices agree on the sign of $R_0-1$, they can lead to different values of $R_0$.  
We refer to \cite{brouwer} for a recent didactic note about this issue.

\medskip
In the context of age-structured models, which are often formulated as integro-partial differential equations with nonlocal boundary conditions, 
reproduction numbers are again associated with the linearization of the model around an equilibrium and the splitting of the linearization into birth and transition; however, in this case, the operators act on infinite-dimensional spaces.  
For example, it is well known that for a susceptible-infected-removed (SIR) model structured by demographic age without vertical transmission, $R_0$ can be computed as the spectral radius of a \emph{Next Generation Operator} (NGO) \cite{diekmann1990}, which is obtained by splitting the linearized operator in two parts--one accounting for all the infection processes and one accounting for all the remaining processes--which are both linear and defined on a subspace of the state space $L^1$, with values in $L^1$ itself \cite[Section 6.2]{Inaba2017}. Thus, the procedure is very similar to that of the NGM method, but considering a space of $L^1$ functions rather than $\R^d$, for $d$ a positive integer. Working in $L^1$ is quite straightforward when processes described by boundary conditions are considered as transition processes. 
However, things get more involved when boundary conditions involve birth/infection processes (for instance, think at vertical transmission, or simple infection in the case of models structured by infection age). In this context, different strategies have been proposed in the literature. In \cite{ Barril2021, barril2018practical}, the authors introduced sequences of ``approximating problems'', for which the relevant reproduction numbers can be computed via the NGO method in the $L^1$ framework and such that these ``approximating''  reproduction numbers converge to the one of the original problem, see \cite{Barril2023} for its applications. Alternatively, another possible approach is to work in extended spaces of the form $\R^d\times L^1$ and to develop the spectral theory following the results of \cite{Thieme2009}; see \cite{Thieme1990} for details on the extended space method and \cite{Inaba2017, Inaba2006} for applications in this context. 

\medskip  
Since reproduction numbers for age-structured models are defined as the spectral radius of operators acting between infinite-dimensional vector spaces, their analytical computation is typically difficult, unless one makes additional simplifying assumptions on the model coefficients (e.g., separable mixing). To overcome this problem, several numerical methods have been proposed to approximate $R_0$ by discretizing the birth and transition operators first to derive a finite-dimensional approximation of the NGO. The NGO is positive and, typically, compact, hence its spectral radius is a dominant eigenvalue (in the sense of largest in magnitude) \cite{KreinRutman1948} and the spectral radius of the discrete operator gives an approximation of $R_0$. \cite{Theta2019, Kuniya2017} proposed a Theta method and a backward Euler method, respectively, to discretize the birth and the transition operators relevant to an age-structured epidemic model with no vertical transmission and a finite age span. Being based on finite-order methods, these two approaches guarantee, under suitable smoothness assumptions on the coefficients, a finite order of convergence. An improvement of these methods was proposed in \cite{BredaDeReggiScarabelVermiglioWu2022, BredaFlorianRipollVermiglio2021, BredaKuniyaRipollVermiglio2020} by using pseudospectral collocation (thus potentially guaranteeing an infinite order of convergence for smooth coefficients \cite{Trefethen2000}) in the case of models with nontrivial boundary conditions and a finite age span.
However, all these methods rely on the definition of $R_0$ in the $L^1$ framework discussed above, thus including the boundary condition in the domain of the transition operator and suffering from a lack of flexibility in the choice of the birth and transition processes. 

\medskip 
In this paper, we introduce a general numerical method to approximate the reproduction numbers of a large class of multi-group age-structured models. We follow the idea of \cite{Theta2019, Kuniya2017, BredaDeReggiScarabelVermiglioWu2022, BredaFlorianRipollVermiglio2021, BredaKuniyaRipollVermiglio2020} by identifying a birth and a transition operator and discretizing them via pseudospectral collocation.
To work with complete flexibility in the definition of the birth/infection and transition processes, we build our approach on the extended space framework by \cite{Inaba2017, Inaba2006}. We define a suitable integral mapping from the extended space to the space of absolutely continuous functions, so that point evaluation is well defined and the birth processes included in the boundary conditions become part of the action of a new operator with a trivial boundary condition. 
The idea of going to the integrated framework has been previously successfully applied in \cite{ando2023, BredaDereggiVermiglio2023, ScarabelBredaDiekmannGyllenbergVermiglio2021, ScarabelDiekmannVermiglio2021}.

We assume that the maximum age is finite, which is equivalent to require that the survival probability (i.e., the probability of still being alive or infectious depending on the context) is zero after the maximum age.
We focus on the applications of the method and we postpone the proof of convergence to a manuscript in preparation by the authors \cite{DereggiScarabelVermiglio}.

\medskip 
The paper is organized as follows. In \Cref{s3}, we consider a prototype linear multi-group age-structured model and, with reference to it, we illustrate the theoretical framework and the reformulation via integration of the state in the age variable. The numerical method is described in \Cref{SectionNumerics}, alongside additional details about its implementation. In \Cref{sec:examples}, we present some applications of the method to epidemic models taken from the literature. To illustrate the flexibility of the approach, we compute different types of reproduction numbers, depending on different interpretations of the age variable and the transmission term. To facilitate the reading, the modeling details and specific computations are collected in \Cref{app:models}. Finally, in \Cref{end} we discuss some concluding remarks.

\section{A general theoretical framework}\label{s3}

In this section, we introduce a general, linear, multi-group, age-structured, population model, which encompasses many models of the literature typically obtained from the linearization of a nonlinear model around an equilibrium. With reference to this prototype model and following \cite{Inaba2017, Inaba2006, Inaba2012}, we first recall the definition of the basic reproduction number and other relevant reproduction numbers useful to address some control strategies of infectious diseases within the extended space framework. Then, we introduce an integral mapping to the space of absolutely continuous functions and provide the equivalent definitions within the $AC$-framework, which is more advantageous for the development of the discretization technique presented in \Cref{SectionNumerics}. 

\medskip
Let $a^{\dagger}\in (0, +\infty)$ denote the maximum age. We consider the following linear, multi-group, age-structured, population model:
\begin{numcases}{}
\partial_t x(t, a)+\partial_a x(t, a)= \displaystyle{\int_0^{a^\dagger}}\beta (a, \xi) x(t,\xi) \dd \xi +\delta(a) x(t,a), \label{prototype-PDE}\\ 
x(t, 0)=\displaystyle\int_0^{a^\dagger}b(a) x(t, a)\dd a, \label{prototype-BC}
\end{numcases}
where $x(t, \cdot) \in X:=L^1([0, a^\dagger], \R^d)$ for $t\ge 0$, \ $\beta \in L^\infty([0,a^\dagger]^2,\R^{d\times d})$, $b,\delta \in L^\infty([0,a^\dagger],\R^{d\times d})$, and $d$ is a positive integer. The $d \times d$ matrices $\beta(a, \xi)$ and $b(a)$ are assumed to be non-negative. 
The $d \times d$ matrix $\delta(a)$ has non-positive diagonal elements and all the off-diagonal elements are non-negative. Hence, $\delta(a)$ is an essentially non-negative matrix and the associated fundamental solution matrix is a non-negative, non-singular matrix \cite[pag.~77]{Inaba2017}.

In the context of infectious disease modeling, \eqref{prototype-PDE}--\eqref{prototype-BC} enables one to describe several types of transmission routes and biological processes: the boundary term \eqref{prototype-BC} can account for either vertical transmission (when $a$ represents demographic age) or for natural infection (when $a$ represents infection age), while the right-hand side of \eqref{prototype-PDE} includes horizontal transmission terms, as well as the removal of individuals via death or recovery. 

To allow for flexibility in the definition of reproduction numbers associated to different ways of inflow into the infected compartments, we assume that $\beta$ and $b$ can be divided into the following:
\begin{equation*}\label{split}
 \beta = \beta^{+} + \beta^{-}, \qquad b = b^{+} + b^{-},
\end{equation*}
where the non-negative matrices $\beta^{+}, \beta^{-}, b^{+}, b^{-}$ are chosen to conveniently split the inflow processes into two parts where $\beta^+, b^+$ collect the birth processes and $\beta^-, b^-$ collect the transition processes.  The $+/-$ notation was inspired by \cite{Inaba2008}.

\medskip
We work in the extended state space of the density function and its boundary value, hence we consider the space $Z:= \mathbb{R}^d \times X$, equipped with the following norm: 
$$\|(\alpha;\phi)\|_Z:=|\alpha|+\|\phi\|_X,$$
where $|\cdot|$ is a norm in $\mathbb{R}^d$ and $\| \cdot \|_X$ is the standard $L^1$ norm, see \cite{Inaba2006} and \cite[Chapter 6.4.2]{Inaba2017}. Furthermore, we introduce the subspace $Z_0 = \{ 0 \} \times X \subset Z$, where $0$ is now used to denote the null vector of $\mathbb{R}^d$.

We define the baseline transition operator $\mathcal{M}^Z\colon D(\mathcal{M}^Z)(\subset Z_0) \to Z$ as follows:
\begin{equation*}\label{Acheserve}
\mathcal{M}^Z (0;\phi):=(\phi(0);\ \phi'-\delta\phi),
\qquad 
(0;\phi) \in D(\mathcal{M}^Z):=\left\{(0; \phi) \in Z\ |\ \phi'\in X\right\},   
\end{equation*}
and the bounded linear birth operators $\mathcal{B}^Z_{\pm} \colon Z_0 \to Z$ such that
\begin{equation*}
\mathcal{B}^Z_{\pm}(0;\phi):=\left(\int_0^{a^\dagger} b^{\pm}(\xi) \phi(\xi)\dd \xi;\int_0^{a^\dagger} \beta^{\pm}(\cdot, \xi) \phi(\xi) \dd \xi \right).
\end{equation*}

From the assumption on $\delta,$ it follows that $\mathcal{M}^Z$ is invertible, and then we can define the following:
$$\mathcal{K}^Z_{\pm}:= \mathcal{B}_{\pm}^Z(\mathcal{M}^Z)^{-1}.$$

To compute the reproduction number for the birth process described by $\mathcal{B}^Z_+$, we define the transition operator $\mathcal{M}_{-}^Z:=\mathcal{M}^Z-\mathcal{B}^Z_{-},$ whose domain coincides with $D(\mathcal{M}^Z).$ 
From $\mathcal{M}_{-}^Z=(\mathcal I-\mathcal{K}^Z_{-})\mathcal{M}^Z$, we have that if $\rho(\mathcal K_{-}^Z)<1$, then $\mathcal{M}_{-}^Z$ is invertible \cite[Section 4]{Inaba2012}. 

Then, the \emph{reproduction number} $R$ for the birth process $\mathcal{B}^Z_{+}$ and the transition operator $\mathcal{M}_{-}^Z$ is the spectral radius of the positive operator 
\begin{equation} \label{RO}
\mathcal H^Z:= \mathcal{B}^Z_{+} (\mathcal{M}_{-}^Z)^{-1}=\mathcal{K}^Z_{+}(\mathcal I- \mathcal{K}^Z_{-})^{-1}. 
\end{equation}
If \eqref{RO} is a compact operator with positive spectral radius, then the reproduction number is a dominant real eigenvalue with an associated non-negative eigenfunction \cite{KreinRutman1948}.
Note that, in line with the interpretation, the operator $(\mathcal I-\mathcal K^Z_-)^{-1} = \sum_{i=0}^\infty (\mathcal K^Z_-)^i$ captures the chains of transmission through any sequence of any other type not included in $\mathcal{B}_+^Z$. 

We generically refer to ``reproduction number'' to include several different interpretations as specific cases, including the basic reproduction number $R_0$ and the type reproduction number $T$, as well as more general definitions \cite{HEESTERBEEK20073,Inaba2017, driessche2017}.
\begin{itemize}
    \item If $\mathcal{B}^Z_+$ contains all processes leading to new infections, then \eqref{RO} is the standard NGO and its spectral radius is precisely $R_0$.
    \item If $\mathcal{B}^Z_+$ only contains a subset of the new infections (e.g., horizontal vs vertical, vector vs host) and $\mathcal{B}^Z_-$ contains all other processes, then \eqref{RO} is the \emph{type reproduction operator}
    and its spectral radius is the type reproduction number, and $\mathcal{K}^Z_+$ and $\mathcal{K}^Z_-$ are the \emph{type-specific NGOs} \cite[pag. 477]{Inaba2017}.
    \item If $\mathcal{B}^Z_+$ contains the inflow in a generic compartment (possibly not a state-at-infection), then the spectral radius of \eqref{RO} is the \emph{state reproduction number} according to the terminology in \cite{Inaba2008}. 
\end{itemize}
Moreover, depending on the assumptions on population immunity and interventions, \eqref{RO} can capture the concepts of \emph{effective} and \emph{control} reproduction numbers. 
Additionally, \eqref{RO} provides a unifying abstract framework to introduce the numerical method in Section \ref{SectionNumerics}.

\medskip
Now, let us consider the Banach space $Y:=AC([0, a^\dagger],\R^d)$ equipped with the norm
\begin{equation*}
\|\psi\|_Y:= |\psi(0)|+\|\psi'\|_X,
\end{equation*}
and its closed subspace $Y_0=\{\psi\in Y\ |\ \psi(0)=0\}$. The integral operator $\mathcal V\colon Z\to Y$ given by
 \begin{equation*}\label{isom}
\mathcal V(\alpha;\phi):=\alpha+\int_0^\cdot \phi(\xi)\dd \xi, 
\end{equation*}
defines an isomorphism between $Z$ and $Y$ and between $Z_0$ and $Y_0$.

By defining $y(t,\cdot):= \mathcal V(0;x(t,\cdot))$, the model \eqref{prototype-PDE}--\eqref{prototype-BC} for $x(t,\cdot) \in X$ is equivalent to the following multi-group model:
\begin{equation}\label{inteq}
\begin{cases}
\partial_t y(t, a)+\partial_a y(t, a)= \displaystyle\int_0^{a^\dagger}b(\xi) y(t, \dd\xi) + \displaystyle\int_0^a\int_0^{a^\dagger}\beta(\zeta, \xi) y(t, \dd\xi)\dd\zeta
+\int_0^a\delta(\xi) y(t,\dd\xi),\\[3mm]
y(t, 0)=0,\\
\end{cases}
\end{equation}
for $y(t,\cdot) \in Y_0$. Note that, in \eqref{inteq}, we consider an absolutely continuous function as a measure. Now, we introduce the birth operators $\mathcal{B}^Y_{\pm} \colon Y_0\to Y$ 
\begin{equation*} \label{conjugationB}
\mathcal{B}^Y_{\pm} := \mathcal{V} \mathcal{B}^Z_{\pm} \mathcal{V}^{-1},    
\end{equation*}
and the transition operators $\mathcal{M}^Y,\mathcal{M}_{-}^Y:D(\mathcal M^Y) (\subset Y_0) \to Y$ 
\begin{equation*} \label{conjugation}
    \mathcal{M}^Y := \mathcal{V} \mathcal{M}^Z \mathcal{V}^{-1}, \quad \mathcal{M}_{-}^Y := \mathcal{V} \mathcal{M}_{-}^Z \mathcal{V}^{-1},
\end{equation*}
where $D(\mathcal M^Y) := \mathcal{V}(D(\mathcal M^Z)).$  Explicitly, they read as
\begin{align*}
\mathcal B^Y_{\pm}\psi&:=\int_0^\cdot\int_0^{a^\dagger}\beta^{\pm}(\zeta, \xi )\psi'(\xi)\dd\xi\dd\zeta+\int_0^{a^\dagger}b^{\pm}(\xi)\psi'(\xi)\dd\xi,  \qquad \psi \in Y_0,\\
\mathcal M^Y\psi&:=\psi'-\int_0^\cdot \delta(\xi)\psi'(\xi)\dd\xi,\qquad \psi\in D(\mathcal M^Y):=\{\psi \in Y_0\ |\ \psi'\in Y\},
\end{align*}
and 
\begin{equation*}
\mathcal{M}_{-}^Y= \mathcal M^Y-\mathcal B_{-}^Y.
\end{equation*}

It is clear that the spectral radius of $\mathcal{H}^Y:=\mathcal{V}\mathcal{H}^Z\mathcal{V}^{-1}$ coincides with $R$. In fact, the relations $\sigma(\mathcal H^Z)=\sigma(\mathcal H^Y)$ \cite[Section 2.1]{EngelNagel2000} and $\sigma_p(\mathcal H^Z)=\sigma_p(\mathcal H^Y)$ \cite[Proposition 4.1]{bredaeig2012} hold, and there is a one-to-one correspondence of the eigenfunctions via the operator $\mathcal{V}$. Moreover, the compactness results for $\mathcal{H}^Z$ in $Z$ can be easily extended to the corresponding operators in $Y$ via the isomorphism $\mathcal{V}.$

\section{Numerical approximation via pseudospectral method}\label{SectionNumerics}
In this section, we illustrate how to approximate the reproduction number of a class of models that can be recast in the framework of \Cref{s3}. The idea is to derive a finite-dimensional approximation $\mathcal H_N^Y$ of $\mathcal H^Y$, and to approximate the eigenvalues of the latter through those of the former. This is achieved by separately discretizing the operators $\mathcal B_+^Y$ and $\mathcal M_-^Y$ in \Cref{s3} via pseudospectral collocation \cite{Trefethen2000, Boyd2001}.

In this section, we only work in the space $Y$ and, to simplify the notation, we drop the superscript $Y$ from the operators acting on $Y$. We adopt the MATLAB-like notation according to which elements of a column vector are separated by ``;'', while elements of a row vector are separated by ``,''.

\medskip
Let us consider the space $Y_N$ of algebraic polynomials of degree at most equal to $N$, for $N$ a positive integer, on $[0, a^\dagger]$, taking values in $\R^d$,  and its subspace $Y_{0,N}:=\{\psi_N\in Y_N\ |\ \psi_N(0)=0\}$. 
For the Chebyshev zeros $\Theta_N:=\{a_1<\dots <a_N\}$ in $(0, a^\dagger)$ \cite{Trefethen2013}, we introduce the restriction operator
$\mathcal R_N\colon Y\to \R^{dN}$ defined as
\begin{equation*}
\mathcal R_N\psi:=(\psi(a_1);\dots; \psi(a_N)), \quad \psi \in Y,
\end{equation*}
and the prolongation operator $\mathcal P_{0,N}\colon \R^{dN}\to Y_{0,N}\subseteq Y_0$ defined as\footnote{Observe that $\Psi_i\in\R^d$ for every $i=1,\dots, N$.} 
\begin{equation*}
\mathcal P_{0,N}(\Psi_1;\dots; \Psi_N):=\sum_{i=1}^N\ell_{0,i}\Psi_i, \quad (\Psi_1;\dots;\Psi_N) \in \mathbb{R}^{dN},
\end{equation*}
for $\{\ell_{0,i}\}_{i=0}^N$ the Lagrange basis relevant to $\Theta_{0,N}:=\{a_0= 0\}\cup \Theta_N$, i.e.,
\begin{equation*}
\ell_{0,i}(a):=\prod_{\substack{j=0\\ i\ne j}}^N \cfrac{a-a_j}{a_i-a_j},\quad a\in[0, a^\dagger],\ \ i=0,\dots, N.
\end{equation*}
Observe that $\mathcal R_N\mathcal P_{0,N}=\mathcal I_{\R^{dN}}$ and that the composition
\begin{equation*}
\mathcal P_{0,N}\mathcal R_{N} = \mathcal L_{0,N}
\end{equation*}
defines the Lagrange interpolation operator $\mathcal L_{0,N}\colon Y_0\to Y_{0,N}$ relevant to $\Theta_{0,N}$.

\medskip
We derive the finite-dimensional approximations $\mathcal B_{N}\colon \R^{dN}\to \R^{dN}$ and $\mathcal M_{N}\colon \R^{dN}\to \R^{dN}$ of $\mathcal B_+$ and $\mathcal M_-$, respectively, as  follows:
\begin{equation*}
    \mathcal B_{N}:=\mathcal R_{N}\mathcal B_+ \mathcal P_{0,N}, \qquad \mathcal M_{N}:=\mathcal R_{N}\mathcal M_- \mathcal P_{0,N}.
\end{equation*}

Then, the finite-dimensional approximation $\mathcal H_N\colon\R^{dN}\to \R^{dN}$ of $\mathcal H$ is obtained as $\mathcal H_N:=\mathcal B_N\mathcal M_N^{-1}$. Finally, we can use the eigenvalues of $\mathcal H_N$ to approximate those of $\mathcal H$. 
The discrete eigenvalue problem can be solved either by using the standard MATLAB function \texttt{eig} or by solving the generalized eigenvalue problem $\mathcal B_{N}=\lambda \mathcal M_{N}$ \cite{BredaFlorianRipollVermiglio2021}. Correspondingly, observe that the eigenvectors of $\mathcal H_N$ give an approximation of the values of the eigenfunctions of $\mathcal H$ at the Chebyshev zeros. Thus, an  approximation of the eigenfunctions of $\mathcal H$ can be obtained by interpolating the eigenvectors of $\mathcal H_N$ at the nodes in $\Theta_{N}$ and, subsequently, an approximation of the eigenfunctions of $\mathcal H^Z$ can be obtained by applying $\mathcal V^{-1}$ to those polynomials.

\subsection{Implementation issues}
Here, for the sake of simplicity, we restrict to the case $d=1$, and we give an explicit description of the entries of the matrices representing the discretized operators.
These follow from the following cardinal property of the Lagrange polynomials:
\begin{equation*}
    \ell_{0,j}(a_i)=\begin{cases}
     1\quad &\text{ if }i=j,\\
     0&\text{otherwise},
    \end{cases}\quad i,j=0,\dots, N,
\end{equation*}
from which it is easy to see that the entries of the matrices $\mathcal B_{N}$ and $\mathcal M_{N}$ are explicitly given by
\begin{equation}\label{B_N_matrix}
(\mathcal B_{N})_{ij}=\int_0^{a_i}\int_0^{a^\dagger}\beta^+(\zeta, \xi)\ell_{0,j}'(\xi)\dd \xi\dd \zeta+\int_0^{a^\dagger}b^+(\xi)\ell_{0,j}'(\xi)\dd \xi,\qquad
i,j=1,\dots N,    
\end{equation} 
and
\begin{align}\label{C_N_matrix}
\notag(\mathcal M_{N})_{ij}=\ &\ell_{0,j}'(a_i)-\int_0^{a_i}\delta(\xi)\ell_{0,j}'(\xi)\dd \xi\\&-\int_0^{a_i}\int_0^{a^\dagger}\beta^-(\zeta, \xi)\ell_{0,j}'(\xi)\dd \xi\dd \zeta+\int_0^{a^\dagger}b^-(\xi)\ell_{0,j}'(\xi)\dd \xi,\qquad
i,j=1,\dots N.    
\end{align}

\medskip
If the integrals in \eqref{B_N_matrix} and \eqref{C_N_matrix} cannot be analytically computed, then we need to approximate them via a quadrature formula. In this regard, for a function $\psi\colon [0, a^\dagger]\to \R^d$, we make the following approximation:
\begin{equation*}
    \int_0^{a^\dagger}\psi(a)\dd a\approx (w_1,\dots, w_N)(
    \psi(a_1);\dots;\psi(a_N)),
\end{equation*}
where $w_1,\dots, w_N$ are the Fejer's first rule-quadrature weights relevant to the Chebyshev zeros \cite{ChebZeros}. As for the integrals in $[0, a_i]$, $i=1,\dots, N$, inspired by \cite{DiekmannScarabelVermiglio2020}, we use the $i$-th row of the inverse of the following (reduced) differentiation matrix:
\begin{equation*}
    (\mathcal D_N)_{ij}:=\ell_{0,j}'(a_i),\quad i,j=1,\dots, N.
    \end{equation*}  

\medskip
When dealing with models where the structuring variable $a$ lives in a ``large'' domain, it can be convenient to consider the change of variable $\alpha=a/a^\dagger$, where $\alpha\in[0, 1]$. Then, by defining $u(t,\alpha):=x(t, a^\dagger \alpha)$, \eqref{prototype-PDE}-\eqref{prototype-BC} can be rewritten as follows:
\begin{equation*}
\begin{cases}
\partial_t u(t, \alpha)+\cfrac{1}{a^{\dagger}}\cdot \partial_\alpha u(t, \alpha)= \displaystyle{\int_0^1} a^{\dagger} \beta(a^\dagger\alpha, a^\dagger \theta) u(t,\theta) \dd \theta +\delta(a^\dagger\alpha) u\left(t,\alpha\right),\\
\cfrac{1}{a^\dagger}\cdot u(t, 0)=\displaystyle\int_0^{1}b(a^\dagger\alpha) u(t, \alpha)\dd \alpha.
\end{cases}
\end{equation*}\smallskip
Via integration, one can derive the corresponding equation for \eqref{inteq}:
\begin{align*}
\partial_t v(t, \alpha)+\frac{1}{a^\dagger}\partial_\alpha v(t, \alpha)=&\displaystyle\int_0^{1}b(a^\dagger\theta) v(t, \dd\theta)\\ &+ \displaystyle\int_0^\alpha\int_0^{1}a^\dagger \beta(a^\dagger\eta, a^\dagger\theta) v(t, \dd\theta)\dd\eta
+\int_0^1\delta(a^\dagger \theta) v(t,\dd\theta).
\end{align*}
Then, the numerical approach can be easily adapted.
Moreover, observe that one could also be interested in using different interpretations of ``age'' in the same model; for example see \eqref{asymptomatic} which considers both infection age and disease age, or \cite[Section 2.1]{Inaba2016}, where an SIR model was proposed with susceptible individuals structured by the demographic age, infected individuals structured by the infection age, and removed individuals structured by the recovery age.  In this context, different interpretations of the age variable can require one to work with different age intervals (or, equivalently, different maximum ages). The method can be easily extended to account for this case by resorting to the scaling described above (with appropriate modifications).

\medskip
Finally, in the presence of breaking points (i.e., discontinuities in the model coefficients or in their derivatives), it is preferable to resort to a piecewise approach. More in the detail, given $0=\bar a_0<\bar a_1<\dots<\bar a_M=a^\dagger$ the breaking points of the coefficients, for $M$ a positive integer, we approximate a function $\psi\in Y$ via a continuous function $\psi_N$ such that ${\psi_{N}}_{|_{[\bar a_{i-1}, \bar a_i]}}$ is a polynomial of degree at most $N$ on $[\bar a_{i-1}, \bar a_i]$ for every $i=1,\dots, M$. 

In this case, in order to simplify the implementation, a possible choice is to extend the mesh (of Chebyshev zeros plus the left endpoint) within each interval by adding the right endpoint. This still ensures convergence of interpolation under the choice made at the beginning of the section \cite[Theorem 4.2.4]{Mastroianni}. Alternatively, it can be convenient to choose the Chebyshev extremal nodes as discretization points and the Clenshaw--Curtis quadrature formula to approximate the integrals in $[0, a^\dagger]$ \cite{ClenshawCurtis1960, Trefethen2008}. The latter choice has been widely used in the literature for pseudospectral methods, and experimentally shows convergence properties comparable to those of Chebyshev zeros \cite{Trefethen2008}. In the codes available at \url{https://cdlab.uniud.it/software}, we implement this latter choice.

\section{Age-structured epidemic models and reproduction numbers in applications} \label{sec:examples}
In this section, we introduce some examples of linear age-structured models in the context of infectious disease dynamics which are obtained from the linearization of a nonlinear model around an equilibrium.
For each of them, we compute different types of reproduction numbers depending on different interpretations of the age variable and the transmission term.
The examples are chosen to cover a range of cases: the first two models are somewhat simplified, which allows us to work with continuous rates and scalar equations and to have explicit expressions for the reproduction numbers; the third example is a system of equations that is useful to reflect on the interpretation of the birth processes; and finally, the last example involves a system of equations and application to real data, which requires piecewise constant parameters. All the reproduction numbers are computed using the method of \Cref{SectionNumerics} with $N=100$ and the piecewise version of the numerical approach in the presence of breaking points. The modeling details and the linearization around the equilibria are described in the appendix, while \Cref{tab:maxitable} shows how the test examples fit into the general framework of \Cref{s3}.
As we assume to work with a finite maximum age, we implicitly assume that the death/removal rates are infinite after the maximum age.

\medskip
\newcommand{\specialcell}[2][c]{%
  \renewcommand*\arraystretch{1}\begin{tabular}[#1]{@{}c@{}}#2\end{tabular}}
  
\begin{table}[h]
\footnotesize
\begin{center}
{\renewcommand*\arraystretch{1.7}
\begin{tabular}{ccccccccc}
\rowcolor{gray!20}
 Model & $d$ & age & $R$ & $b^+$ & $b^-$ & $\beta^+$ & $\beta^-$ & $\delta$ \\
 \specialrule{\lightrulewidth}{0pt}{0pt}
\eqref{TSI} & 1 & \specialcell{infect.} & $R_c$ & $b$ & 0 & 0 & 0 & $-\gamma$ \\
\rowcolor{gray!10}
 \eqref{multistrain} & 1 & demog. & $R_{0,j}$ & 0 & 0 & $\beta_j$ & 0 & 0 \\
 \rowcolor{gray!10}&    &        & $R^j_k$ & 0 & 0 & $s_k^* \beta_j$ & 0 & 0 \\

\eqref{asymptomatic} & 2 &  \specialcell{infect./\\disease} &  $R_{0}$ & $\begin{pmatrix} b_{11} & b_{12} \\ 0 & 0 \end{pmatrix}$ & $\begin{pmatrix} 0 & 0 \\ b_{21} & 0 \end{pmatrix}$ & 0 & 0 & $\begin{pmatrix} -b_{21}-\gamma_1 & 0 \\ 0 & -\gamma_2 \end{pmatrix}$ \\
   &    &    &   $T_S$ &  $\begin{pmatrix} 0 & 0 \\ b_{21} & 0 \end{pmatrix}$ &  $\begin{pmatrix} b_{11} & b_{12} \\ 0 & 0 \end{pmatrix}$ &  0 &  0 &  $\begin{pmatrix} -b_{21}-\gamma_1 & 0 \\ 0 & -\gamma_2 \end{pmatrix}$ \\
 \cellcolor{gray!10}\eqref{rubella} & \cellcolor{gray!10}2 & \cellcolor{gray!10} demog.  & \cellcolor{gray!10}$R_{0}$ & \cellcolor{gray!10} $\begin{pmatrix} 0 & 0 \\ 0 & b \end{pmatrix}$ & \cellcolor{gray!10} 0 & \cellcolor{gray!10} $\begin{pmatrix} \beta & 0 \\ 0 & 0 \end{pmatrix}$ & \cellcolor{gray!10} 0 & \cellcolor{gray!10} $\begin{pmatrix} -\sigma & 0 \\ \sigma & -\gamma \end{pmatrix}$ \\
 \cellcolor{gray!10} &  \cellcolor{gray!10}  &  \cellcolor{gray!10}  &  \cellcolor{gray!10} $T_H$ & \cellcolor{gray!10} 0 & \cellcolor{gray!10} $\begin{pmatrix} 0 & 0 \\ 0 & b \end{pmatrix}$ & \cellcolor{gray!10} $\begin{pmatrix} \beta & 0 \\ 0 & 0 \end{pmatrix}$ & \cellcolor{gray!10} 0 & \cellcolor{gray!10} $\begin{pmatrix} -\sigma & 0 \\ \sigma & -\gamma \end{pmatrix}$ \\
 
 \cellcolor{gray!10} &  \cellcolor{gray!10}  &  \cellcolor{gray!10}  &  \cellcolor{gray!10} $T_V$ & \cellcolor{gray!10} $\begin{pmatrix} 0 & 0 \\ 0 & b \end{pmatrix}$ & \cellcolor{gray!10} 0 & \cellcolor{gray!10} 0 & \cellcolor{gray!10} $\begin{pmatrix} \beta & 0 \\ 0 & 0 \end{pmatrix}$ & \cellcolor{gray!10} $\begin{pmatrix} -\sigma & 0 \\ \sigma & -\gamma \end{pmatrix}$ \\
 \specialrule{\lightrulewidth}{0pt}{0pt}
\end{tabular}
}

\label{tab:maxitable}
\caption{Birth and transition processes used to compute the reproduction numbers for the models in \Cref{sec:examples}, with reference to the notation used in \Cref{s3}.}
\end{center}
\end{table}

\subsection{An epidemic model structured by infection age} \label{s:tsi}
\begin{table}[ht]
\small
\begin{center}
{\renewcommand*\arraystretch{1.2}
\begin{tabular}{ccl}
\rowcolor{gray!20}
 Symbol & Value & Description \\
 \specialrule{\lightrulewidth}{0pt}{0pt}
$\tau^\dagger$ & 14 & Maximum infection age  (days)\\
\rowcolor{gray!10}
 $b(\tau)$ & $ R_0 \, c \tau^{5.9252}$ & Per capita infection rate at infection age $\tau$  (days$^{-1}$)\\
 $\gamma_0$  & 1.3850 & Per capita baseline recovery rate in $[0,\tau^\dagger]$ (days$^{-1}$) \\
\rowcolor{gray!10}
 $f(\tau)$  & $\Gamma(\mu,\sigma)$ & Incubation period probability density function \\
 $c$ & 0.0152 & Normalizing constant for infectiousness \\
\rowcolor{gray!10}
 $\epsilon$ & varying & Fraction of symptomatic individuals \\
 $D$ & varying & Delay between symptom onset and isolation (days) \\
 \specialrule{\lightrulewidth}{0pt}{0pt}
\end{tabular}
}\smallskip

\label{tab:timesinceinf}
\caption{Parameters of \eqref{TSI}. The function $f$ is a Gamma probability density function with mean $\mu=4.84$ days and standard deviation $\sigma=2.79$, describing the incubation period distribution \cite{Overton}.
The parameters are chosen so that $b(\tau)\mathrm{e}^{-\gamma_0\tau} = R_0 \Gamma(5,1.9)$, where $\Gamma(5,1.9)$ is a Gamma density function with mean $5$ days and standard deviation $1.9$ \cite{Ferretti} (normalized in $[0,\tau^\dagger]$), and $R_0$ is varied in the simulations.}
\end{center}
\end{table}

\begin{figure}
    \centering
    \includegraphics[width=.85\textwidth]{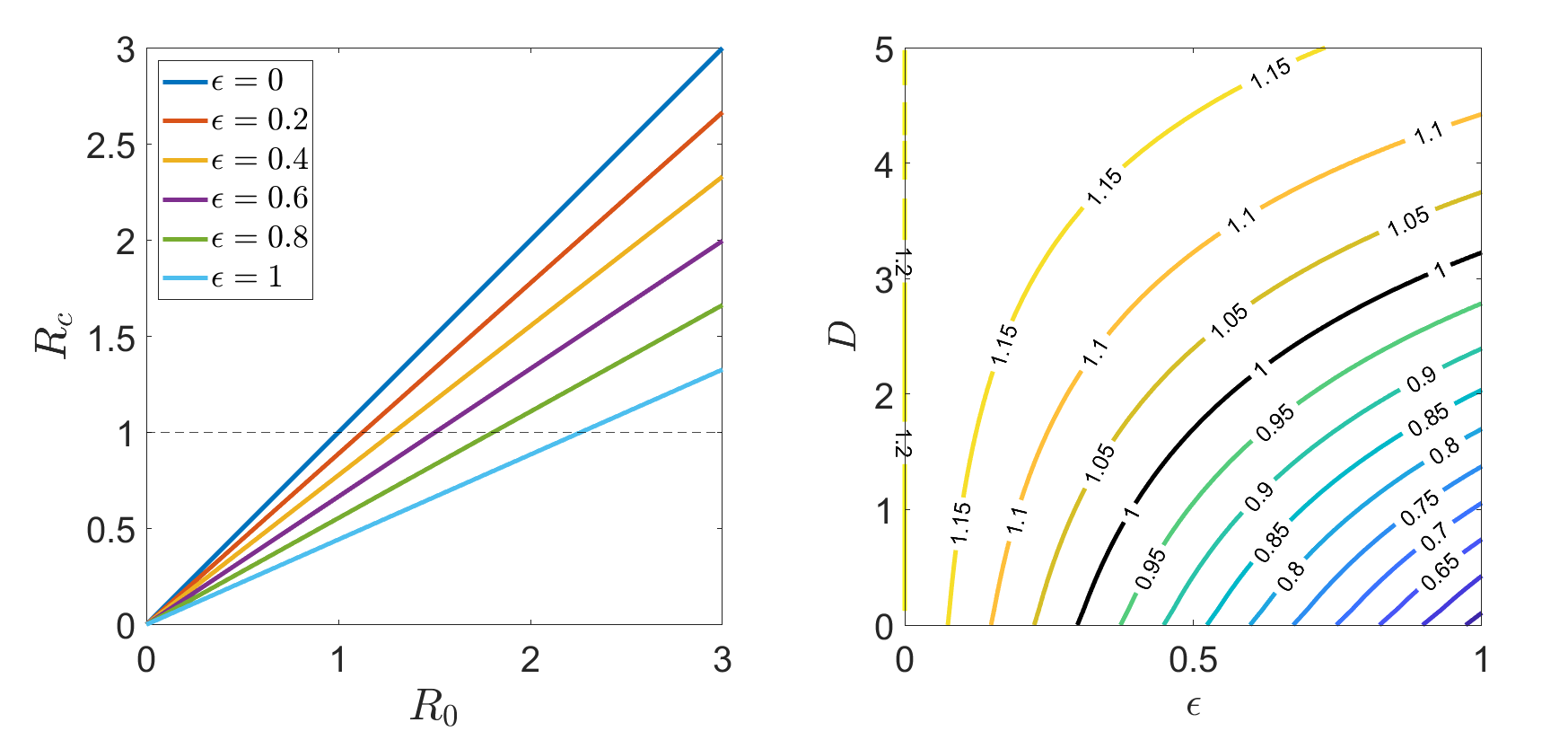}
    \caption{Model \eqref{TSI}. Left: $R_c$ varying the multiplicative parameter $R_0$, for different values of the fraction of symptomatic individuals ($\epsilon$) and assuming no delay from symptoms to diagnosis ($D=0$). Right: $R_c$ varying $\epsilon$ ($x$-axis) and $D$ ($y$-axis), for $R_0 = 1.2$.}
    \label{fig:timesinceinf}
    
    \bigskip   
    
    \bigskip
    
    \bigskip
    \includegraphics[width=0.85\textwidth]{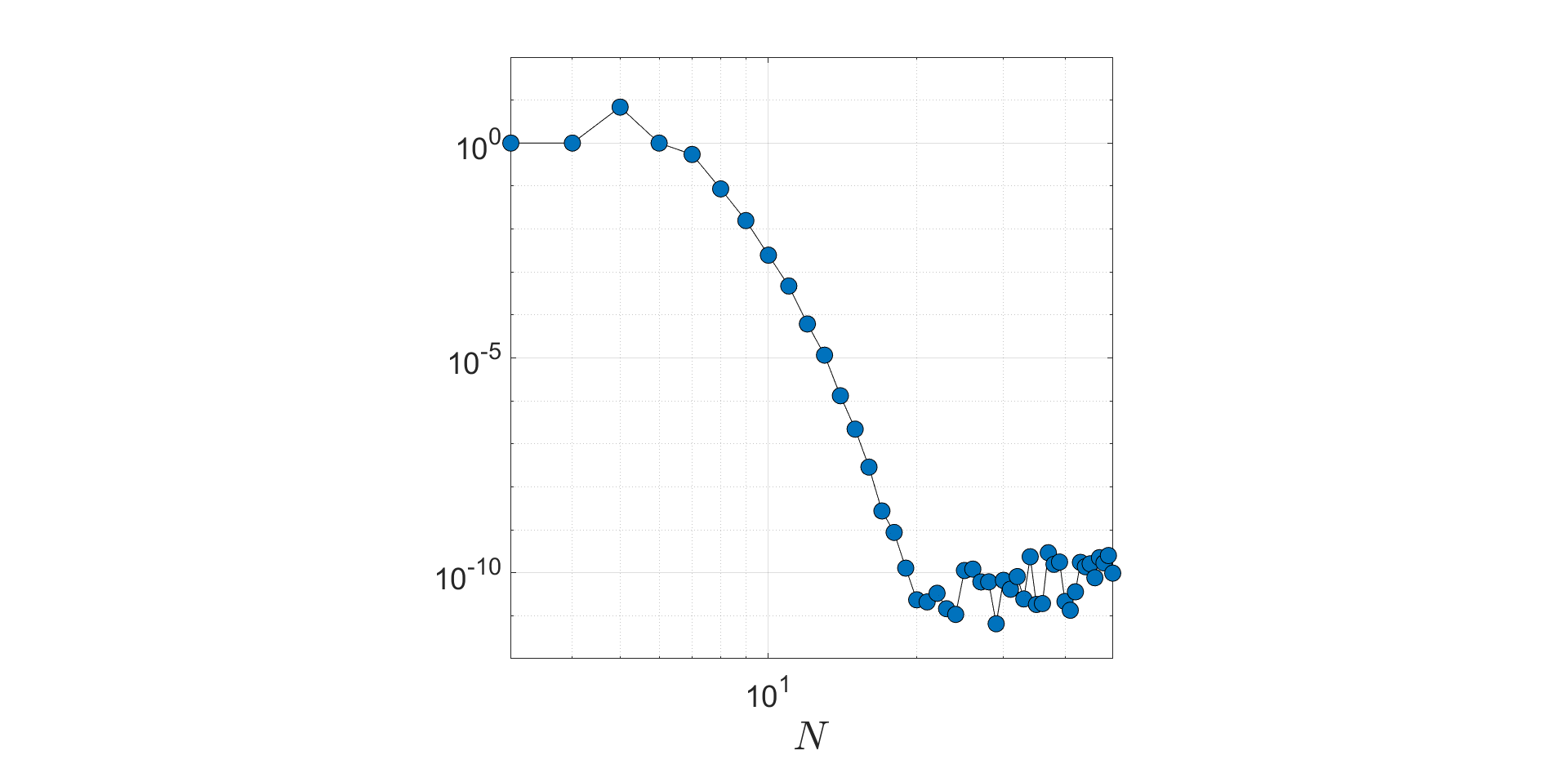}
    \caption{Model \eqref{TSI}. Log-log plot of the absolute error of approximation for $R_0=1$ for increasing $N$ with $\epsilon=0$.}
    \label{fig:timesinceinfconv}
\end{figure}

Let us consider the spread of an infectious disease in a closed population, with the infected individuals structured by infection age, in the presence of isolation measures upon detection of symptoms. We refer to \cite[Section 5.3]{Inaba2017} and \Cref{app:TSI} for further details about the nonlinear model, and to \cite{ScarabelPellisOgdenWu2021} for an application of an extended version of this model to study the impact of contact tracing on the containment of COVID-19.

\medskip
Let $i(t,\tau)$ denote the density of infected individuals at time $t\geq 0$ and infection age $\tau\in [0,\tau^\dagger]$. The linearization around the disease-free equilibrium reads as follows:
\begin{equation}\label{TSI}
\begin{cases}
\partial_t i(t, \tau)+\partial_\tau i(t, \tau)= - \gamma(\tau) i(t,\tau),\\[1mm]
i(t, 0)= \displaystyle\int_0^{\tau^\dagger} b(\tau) i(t,\tau)\dd \tau.
\end{cases}
\end{equation}
The non-negative functions $b,\, \gamma \colon [0,\tau^\dagger] \to\R$ describe the per capita infection rate and recovery rate, respectively. In particular, we assume that $\gamma$ accounts for both natural recovery and isolation upon symptom onset, and we take the following:
\begin{equation*}
    \gamma(\tau) = 
    \begin{cases}
        \gamma_0, &\tau < D, \\
        \gamma_0 + \cfrac{\epsilon f(\tau-D)}{1-\epsilon \int_0^\tau f(\xi-D) \dd \xi}, & \tau \ge D,
    \end{cases}
\end{equation*}
where $\gamma_0,\epsilon,D$ are non-negative parameters whose interpretation is specified in \Cref{tab:timesinceinf}. 

In the absence of isolation from symptoms ($\epsilon=0$), the basic reproduction number is $R_0 = \int_0^{\tau^\dagger} b(\tau) \mathrm{e}^{-\gamma_0 \tau} \dd \tau$. In the presence of isolation ($\epsilon>0$), the control reproduction number is as follows:
\begin{equation*}
R_c = \int_0^{\tau^\dagger} b(\tau) \mathcal{F}(\tau) \dd\tau, \quad \mathcal{F}(\tau) := \mathrm{e}^{-\int_0^\tau \gamma(\theta) \dd \theta}.
\end{equation*}
Note that, even though an explicit expression for $R_c$ is available in this case, its computation from the analytical formula requires numerical approximations.

In \Cref{fig:timesinceinf}, we investigated the impact of a delay from symptom onset to diagnosis ($D$) and the fraction of symptomatic infections ($\epsilon$) using parameter values inspired by the COVID-19 literature, collected in \Cref{tab:timesinceinf}.
$R_c$ increases linearly with the baseline transmission parameter $R_0$, and isolation of symptomatic individuals effectively reduces $R_c$ and promotes controllability (left panel). Moreover, the isolation is more effective if the proportion of symptomatic individuals is larger or the delay from symptoms to isolation is shorter (right).

Finally, in \Cref{fig:timesinceinfconv}, we illustrate the convergence behavior of the approximation error with respect to $R_0=1$ for increasing values of $N$ with $\epsilon=0$.

\subsection{A multi-strain epidemic model with host age structure} \label{sec:multistrain}

\begin{table}[ht]
\small
\begin{center}
{\renewcommand*\arraystretch{1.2}
\begin{tabular}{ccl}
\rowcolor{gray!20}
 Symbol & Value & Description \\
 \specialrule{\lightrulewidth}{0pt}{0pt}
$a^\dagger$ & 20  & Maximum life span (yr) \\
\rowcolor{gray!10}
 $R_0^d$ & 20 & Demographic reproduction number  \\
 $\mu(a)$  & 0.6 & Age-specific death rate in $[0, a^\dagger]$ (yr$^{-1}$)\\
\rowcolor{gray!10}
 $f(a)$ & 0.6 & Age-specific per capita birth rate (yr$^{-1}$) \\
 $\Phi(x)$ & $\frac{1}{1+0.3x}$ & Function describing density dependence of births  \\
\rowcolor{gray!10}
 $K_1(a)$ & $\frac{m_1}{1+c_1a}$ & Age-specific susceptibility of individuals to strain $1$ \\
 $K_2(a)$ & $m_2+c_2a$ & Age-specific susceptibility of individuals to strain $2$ \\
\rowcolor{gray!10}
 $q_j(a)$ & 1 & Age-specific infection rate for strain $j$, $j=1,\,2$ \\
 \specialrule{\lightrulewidth}{0pt}{0pt}
\end{tabular}
}

\label{tab:multistrain}
\caption{Parameters of \eqref{multistrain}, taken from \cite{Qiu2012}. With this parameter choice, the total population density is $\Phi^{-1}(1/R_0^d) = 19/3$. The non-negative parameters $c_1,\, c_2,\, m_1$ and $m_2$ are introduced for mathematical convenience to characterize the shape of the functions $K_1$ and $K_2$, and are varied in the simulations.}
\end{center}
\end{table}

\begin{figure}[pth]
    \centering
    \includegraphics[width=.79\textwidth]{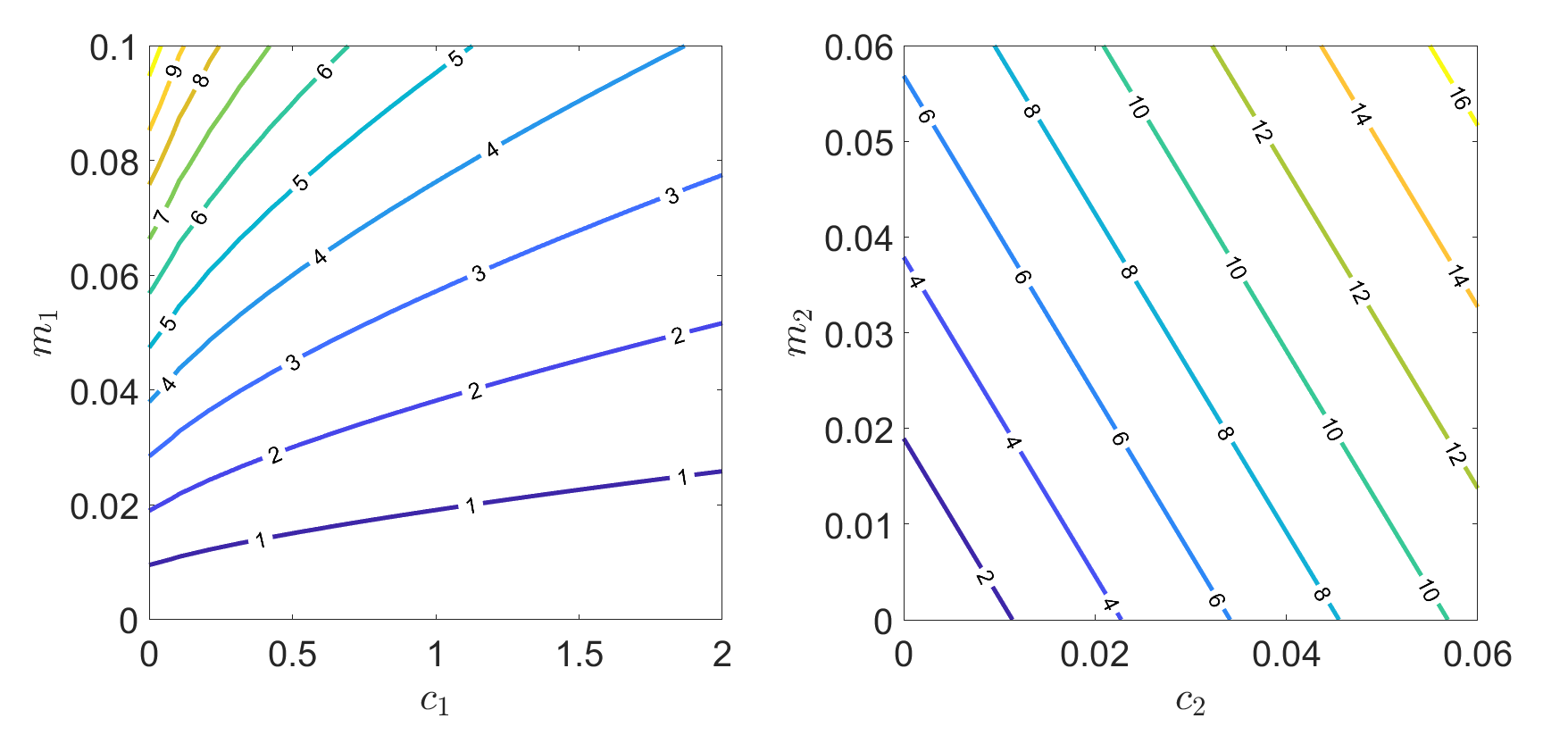}
    \caption{Model \eqref{multistrain}.
    Basic reproduction numbers $R_{0,1}$ (left) and $R_{0,2}$ (right) varying the parameters $c_1, c_2, m_1 $ and $m_2$. When not varied, the parameters are fixed at: $c_1=1$, $ c_2 = 0.06$, $m_1 = 0.1$, and $m_2 = 0.06$.}
    \label{fig:multistrain_R0}\medskip
    \includegraphics[width=.95\textwidth]{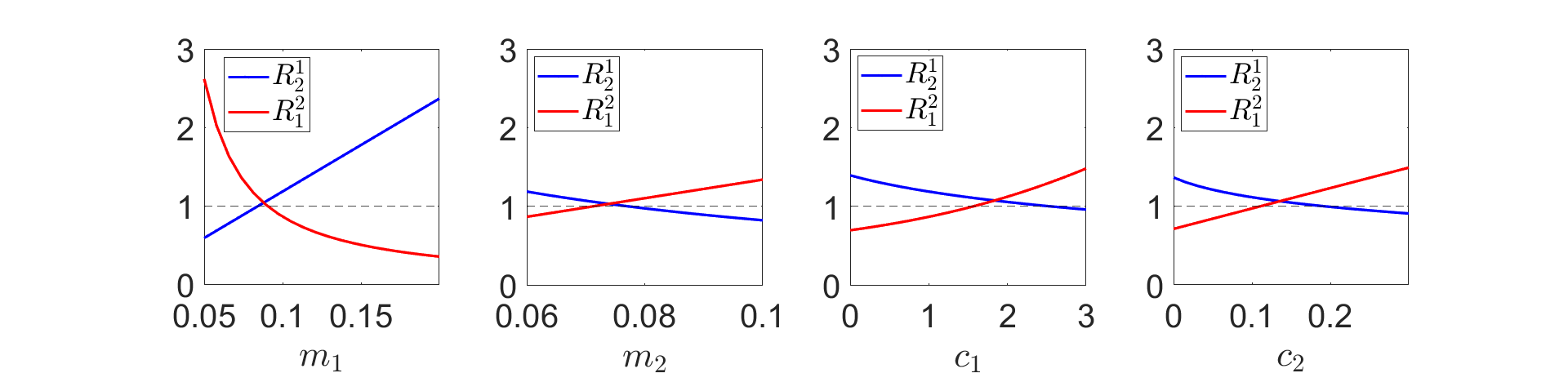}
    \caption{Model \eqref{multistrain}.
    Invasion reproduction numbers varying the parameters $m_1, m_2, c_1 $ and $c_2$. Note that these parameters affect the value of the invasion reproduction numbers both via the kernels and via the value of the susceptible population at equilibrium. When not varied, the parameters are fixed at: $c_1=1$, $ c_2 = 0.06$, $m_1 = 0.1$, and $m_2 = 0.06$. 
    }
    \label{fig:multistrain}
\end{figure}

\begin{figure}
    \centering
    \includegraphics[width=.79\textwidth]{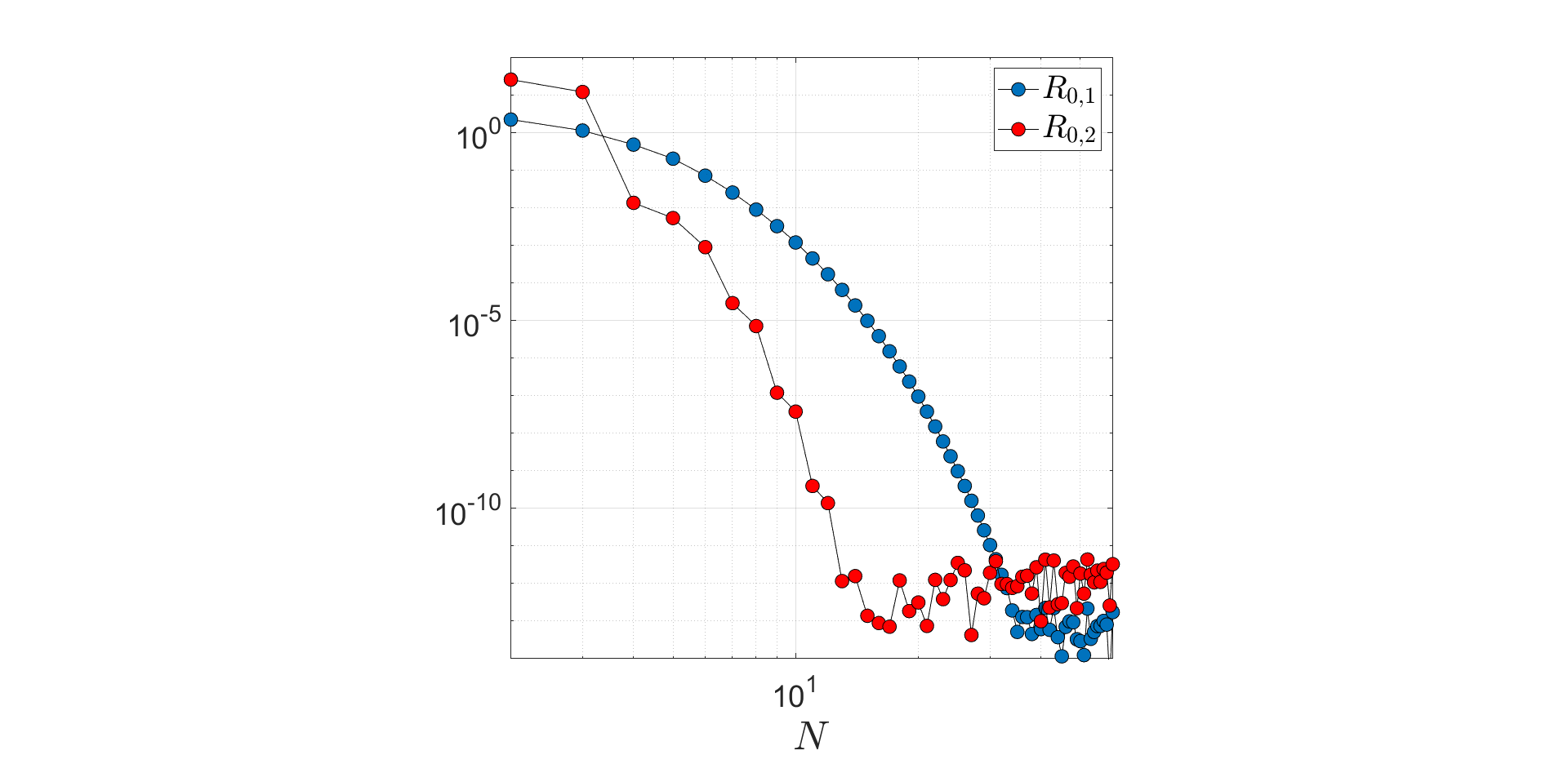}
    \caption{Model \eqref{TSI}. Log-log plot of the absolute error of approximation for $R_{0,1}\approx 5.24$ (blue) and $R_{0,2}\approx 16.88$ (red) for increasing $N$ with $c_1=1,\ c_2=0.06,\ m_1=0.1,\ m_2=0.06$.}
    \label{fig:multistrainconv}
\end{figure}

We consider a model with two classes of infected individuals structured by demographic age, which describes the dynamics of two competing strains in a host population \cite{Qiu2012}. For applications of similar models to real infectious diseases, we refer to \cite{Castillo}, where the case of influenza was discussed. In the model, susceptible individuals can be infected either with strain $1$ or with strain $2$, and enter the class of individuals infected with each strain. Cross-immunity is assumed, so that individuals recovered with any strain are immune towards both strains, and immunity is assumed to be permanent. In \cite{Qiu2012}, it is shown that the coexistence of both strains in an endemic equilibrium is not possible when the parameters do not depend on age, but is possible for age-dependent parameters. 
The model derivation is described in detail in \Cref{app:multistrain}, and the parameters are summarized in \Cref{tab:multistrain}.
We assume that the total population is at the (nontrivial) demographic steady state, and we neglect disease-induced mortality.
The system can have four equilibria, of which three are boundary equilibria (one disease-free and two with only one strain present in the population) and one is  an endemic coexistence equilibrium. 

\medskip
Let $i_j(t,a)$ denote the density of individuals infected with strain $j$, for $j=1,2$, at time $t\geq 0$ and demographic age $a \in [0,a^\dagger]$. The linearization around the disease-free equilibrium reads as follows:
\begin{equation}\label{multistrain}
\begin{cases}
\partial_t i_j(t, a)+\partial_a i_j(t, a)= \displaystyle \int_0^{a^\dagger} \beta_j(a,\xi ) i_j(t,\xi) \dd \xi, \\[3mm]
i_j(t,0)=0,
\end{cases}
\end{equation}
with $\beta_j(a,\xi) = K_j(a) q_j(\xi) P^*(\xi)$, see also \Cref{tab:multistrain}, and 
$$P^*(a) = \frac{\Pi(a)\Phi^{-1}(1/R_0^d)}{\int_0^{a^\dagger} \Pi(\xi) \dd \xi}  , \quad \Pi(a) := \mathrm{e}^{-\int_0^a \mu(\xi) \dd \xi}. $$ 
The basic reproduction number $R_{0,j}$ for strain $j$ can be computed by individually considering each scalar equation in the absence of the other strain, and explicitly reads as follows:
\begin{equation*}
R_{0,j} = \int_0^{a^\dagger} \frac{P^*(a)}{\Pi(a)} \int_a^{a^\dagger} \beta_j(a,\xi)\frac{\Pi(\xi)}{P^*(\xi)} \dd \xi \dd a, \quad j=1,2. 
\end{equation*}
\Cref{fig:multistrain_R0} shows the values of $R_{0,1}$ and $R_{0,2}$ varying the parameters $c_1$, $m_1$, and $c_2$, $m_2$, respectively, which are introduced for mathematical convenience to characterize the shape of the age-specific susceptibility of individuals (see also \Cref{tab:multistrain}). 
The large values of the basic reproduction numbers for these parameter choices ensure the existence of the boundary equilibria where one strain is endemic in the population, and allow us to study the invasion reproduction numbers, as explained below.

Consider the boundary equilibria $E^*_1=(s^*_1,i^*_1,0)$ and $E^*_2=(s^*_2,0,i^*_2)$, where only strain 1 or strain 2, respectively, is present in the population (where $s^*_k(a)$ and $i^*_k(a)$ are the densities of susceptible and infected individuals at equilibrium divided by the stable age distribution $P^*(a)$). The equilibria $E^*_k$, $k=1,2$, do not admit an analytic expression, but their values can be solved numerically, as explained in \Cref{app:multistrain}. 
The linearization at $E^*_k$ for $k=1,2$ and $j\neq k$ reads as follows:
\begin{equation*}
\begin{cases}
\partial_t i_j(t, a)+\partial_a i_j(t, a)= s^*_k(a) \displaystyle \int_0^{a^\dagger} \beta_j(a,\xi) i_j(t,\xi) \dd \xi, \\[3mm]
i_j(t, 0)=0.
\end{cases}
\end{equation*}
The \emph{invasion reproduction number} $R_k^j$, which describes whether strain $j$ can invade the equilibrium set by strain $k$,  admits the following explicit expression:
$$ R^j_k = \int_0^{a^\dagger} s^*_k(a) \frac{P^*(a)}{\Pi(a)} \int_a^{a^\dagger} \beta_j(a,\xi) \frac{\Pi(\xi)}{P^*(\xi)} \dd \xi \dd a.$$
Note that, in this case, computing the reproduction numbers from the analytical formula requires one to numerically approximate not only the integrals, but also the equilibria.

When both invasion reproduction numbers are larger than 1, the coexistence equilibrium is stable and the two strains can persist in the population \cite{Qiu2012}. 

\Cref{fig:multistrain} shows the invasion reproduction numbers $R_2^1$ and $R_1^2$ when varying the parameters $m_1, m_2, c_1$, and $c_2$. As expected, each of these parameters has an opposite impact (either decreasing or increasing) on $R_2^1$ and $R_1^2$. 
The parameter $m_1$ is the only one that positively impacts the invasion of strain 1 (increasing $R_2^1$) and negatively impacts strain 2 (decreasing $R_1^2$). 

In \Cref{fig:multistrainconv}, we plot the absolute errors of approximation for $R_{0,1}$ and $R_{0,2}$ for increasing $N$ with respect to the reference values computed from the analytical formulas. The numerical convergence is of infinite order, which is consistent with the fact that the parameters are of class $C^\infty$.

\subsection{An epidemic model with symptomatic and asymptomatic transmission}
\begin{table}[th]
\small
\begin{center}
{\renewcommand*\arraystretch{1.2}
\begin{tabular}{ccl}
\rowcolor{gray!20}
 Symbol & Value & Description \\
 \specialrule{\lightrulewidth}{0pt}{0pt}
$\tau_1^\dagger$ & 14 & Maximum infection age (days) \\
\rowcolor{gray!10}
$\tau_2^\dagger$ & 14 & Maximum disease age (days) \\
 $\gamma_1$ & 0 & Recovery rate in the asymptomatic phase in $[0,\tau_1^\dagger]$ (days$^{-1}$)\\
\rowcolor{gray!10}
 $\gamma_2$ & 0.45 & Recovery rate in the symptomatic phase in $[0,\tau_2^\dagger]$ (days$^{-1}$)\\
 $b_{21}$ & 0.676 & Rate of developing symptoms (days$^{-1}$) \\
\rowcolor{gray!10}
 $b_{11}$ & varying 
 & Per capita infection rate in the asymptomatic phase (days$^{-1}$) \\
 $ b_{12}$ & 0.0695 & Per capita infection rate in the symptomatic phase (days$^{-1}$)
\end{tabular}
}
\label{tab:asymptomatic_exp}
\caption{Parameters of \eqref{asymptomatic}. Parameter values are taken from \cite{Inaba2008}, assuming exponential distributions (i.e., all rates are assumed to be constant), making exception for $b_{11}$ and $b_{12}$ which are chosen for illustration purposes. }
\end{center}
\end{table}
\begin{figure}[pt]
    \centering
    \includegraphics[width=.8\textwidth]{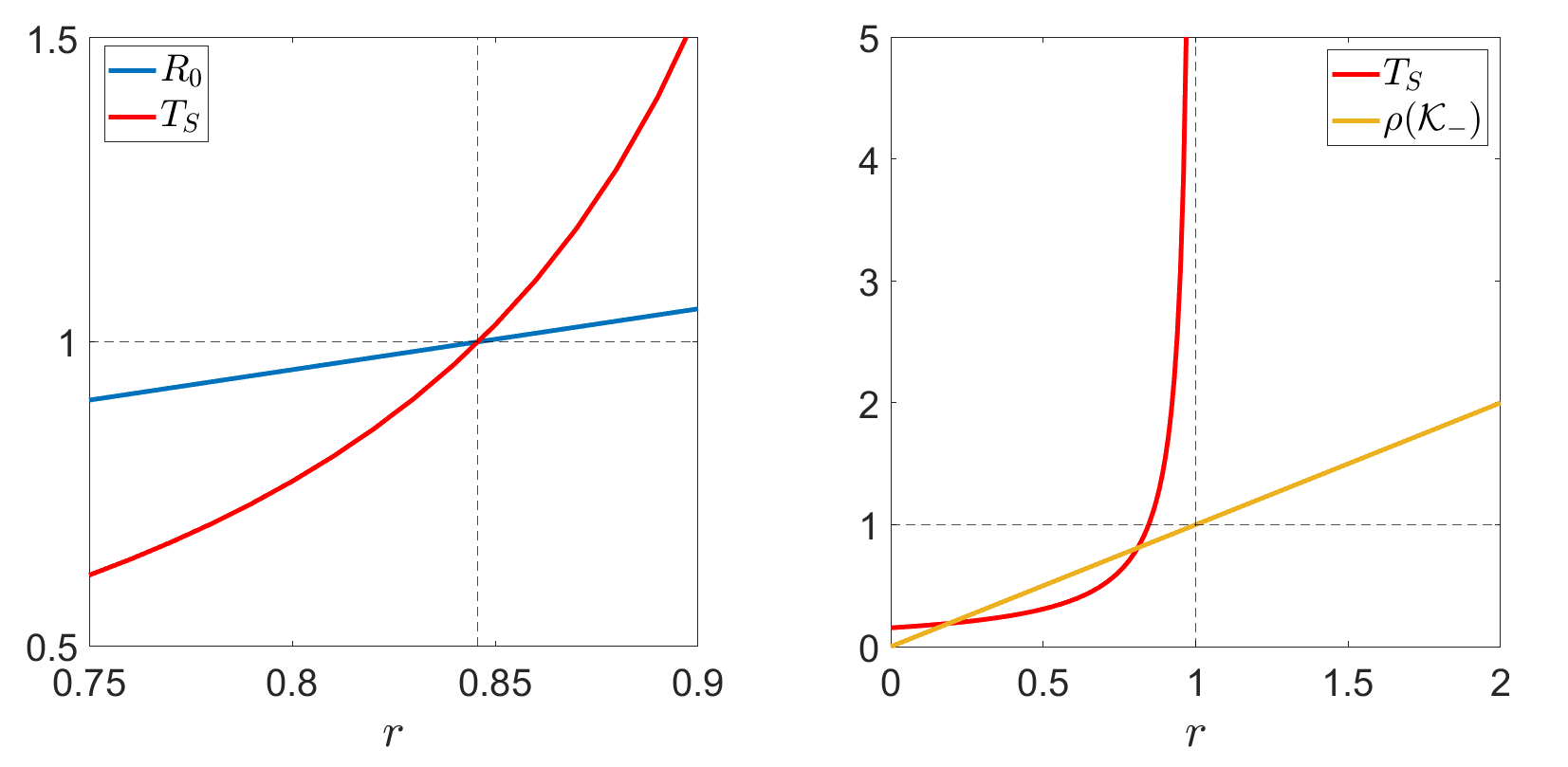}
    \caption{Model \eqref{asymptomatic}. Left: $R_0$ and $T_S$ as functions of $r:= b_{11}/b_{21}$. 
    Right: $T_S$ and the spectral radius of the NGO relevant to the asymptomatic individuals as functions of $r$.}
    \label{fig:sympandasymp}
\end{figure}
We consider the asymptomatic transmission model described in \cite{Inaba2008}. 
Upon infection, individuals enter an asymptomatic phase characterized by an infection age $\tau_1 \in [0,\tau_1^\dagger]$. Then, individuals can either recover without developing symptoms at a rate $\gamma_1(\tau_1)$, or they can develop symptoms at a rate $b_{21}(\tau_1)$, upon which they enter the symptomatic phase, which is characterized by a disease age (time since the onset of symptoms) $\tau_2 \in [0,\tau_2^\dagger]$, and from which they recover with a rate $\gamma_2(\tau_2)$. 

\medskip
Let $i_1(t, \tau_1)$ denote the density of asymptomatic individuals at time $t\ge0$ and infection age $\tau_1$, and let $i_2(t,\tau_2)$ denote the density of symptomatic individuals at time $t\ge0$ and disease age $\tau_2$.
Assuming that the total susceptible population is normalized to one, the linearization around the disease-free equilibrium reads
\begin{equation}\label{asymptomatic}
\begin{cases}
\partial_t i_1(t, \tau_1)+\partial_{\tau_1} i_1(t, \tau_1)=-(b_{21}(\tau_1)+\gamma_1(\tau_1)) i_1(t,\tau_1),\\[3mm]
\partial_t i_2(t, \tau_2)+\partial_{\tau_2} i_2(t, \tau_2)=-\gamma_2(\tau_2) i_2(t,\tau_2),\\[1mm]
i_1(t, 0)=\displaystyle\int_0^{\tau_1^\dagger} b_{11}(\tau_1) i_1(t, \tau_1)\dd\tau_1+\int_0^{\tau_2^\dagger} b_{12}(\tau_2) i_2(t, \tau_2)\dd \tau_2,\\
i_2(t, 0)=\displaystyle\int_0^{\tau_1^\dagger} b_{21}(\tau_1) i_1(t, \tau_1)\dd\tau_1,
\end{cases}
\end{equation}
where $b_{11}(\tau_1)$ is the per capita infection rate at the infection age $\tau_1$ in the asymptomatic phase, and $b_{12}(\tau_2)$ is the infection rate at the disease age $\tau_2$ in the symptomatic phase. 

\medskip
The parameter values used in the numerical simulations are listed in \Cref{tab:asymptomatic_exp}. 
Here, inspired by \cite[Section 4]{didactic} and \cite{Inaba2008}, we can consider different definitions of ``birth''.
According to the standard interpretation of $R_0$ as the number of new infections generated by one average infectious individual, we consider all new infections coming from asymptomatic and symptomatic individuals as birth processes, and consider the development of symptoms as transition process. On the other hand, since asymptomatic individuals are invisible from the point of view of the public health system (in the absence of other interventions like test-and-trace and asymptomatic testing), one could be interested in studying the effectiveness of control measures to the class of symptomatic individuals only (e.g., isolation upon symptoms) \cite{Inaba2008}. In this case, one could interpret the entrance in the class of symptomatic individuals as birth, while the asymptomatic phase is included in the transition operator, see also \cite{Barril2021b}. Therefore, we denote the state reproduction number of symptomatic individuals by $T_S$. As expected, the two reproduction numbers, $R_0$ and $T_S$, have different values in general, but the same threshold at 1, as seen in \Cref{fig:sympandasymp} when varying $r:=b_{11}/b_{21}$.
Additionally, \Cref{fig:sympandasymp} illustrates another important theoretical property: the state reproduction number $T_S$ is finite (and well defined) only when the spectral radius of the NGO relevant to asymptomatic transmission is smaller than one. When the latter becomes larger than one, then asymptomatic individuals alone can sustain the epidemic, hence interventions that are targeted to only symptomatic individuals are not sufficient to control its spread. This feature is reflected in the behavior of $T_S$, which tends to infinity and becomes not well defined.

\subsection{A model for the spread of Rubella with vertical transmission}\label{rubella}

\begin{table}[th]
\small
\begin{center}
{\renewcommand*\arraystretch{1.2}
\begin{tabular}{ccl}
\rowcolor{gray!20}
 Symbol & Value & Description \\
 \specialrule{\lightrulewidth}{0pt}{0pt}
$a^\dagger$ & 75  & Maximum life span (yr) \\
\rowcolor{gray!10}
 $\eta$ & 4 & Rate of loss of protection provided by maternal antibodies (yr$^{-1}$) \\
 $\sigma$  & 34.76 & Rate of acquisition of infectiousness (yr$^{-1}$) \\
\rowcolor{gray!10}
 $\gamma$ & 31.74 & Recovery rate from infectious period (yr$^{-1}$) \\
 $\Pi(a)$ & 1 & Natural survival probability in $[0,a^\dagger]$ \\
\rowcolor{gray!10}
 $f(a)$ & $1/a^\dagger$ & Per capita birth rate (yr$^{-1}$) \\
 $k(a,\xi)$ & See \Cref{tab:rubellaKi}  & Transmission rate (yr$^{-1}$)\\
\rowcolor{gray!10}
 $q$ & 0.9 & Proportion of vertically infected newborns \\
 $\nu$ & varying & Per capita vaccination rate (yr$^{-1}$) \\
 \specialrule{\lightrulewidth}{0pt}{0pt}
\end{tabular}
}
\label{tab:rubella}
\caption{Parameters of \eqref{rubella_lin} taken from \cite{rubella1985}. The functions $f$ and $\Pi$ are chosen so that $\int_0^{a^\dagger} f(a) \Pi(a) \dd a = 1$. }
\end{center}
\end{table}

Rubella, also known as German measles or three-day measles, is an acute and contagious viral infection that can be vertically transmitted \cite{rubellaCDC}. It is not particularly severe in children and adults, but infection during pregnancy can result in the so called congenital rubella syndrome, which can result in fetal death or congenital malformations in infants \cite{WHOrubella}. For women infected during early pregnancy (first trimester), the probability of passing the virus to the fetus is reported to be 90\% \cite{WHOrubella} and, since there is no treatment for Rubella, the design of vaccination policies plays a fundamental role. In the last few decades, this has triggered a series of works by Anderson and colleagues, see for example \cite{rubella1985, rubella1983, Anderson1985}.

\medskip
Here, we consider a model inspired by \cite{rubella1985} for the spread of congenital rubella syndrome in the United Kingdom.
The model definition and the derivation of the disease-free equilibrium and the corresponding linearization are described in detail in \Cref{app:rubella}. The parameter definitions and values used in the numerical tests are collected in \Cref{tab:rubella}. Note that, in the literature, the term control reproduction number is typically used in the presence of interventions such as vaccinations. To simplify our terminology, here we refer to $R_0$ even in the presence of vaccinations. 

\medskip
Let $e(t,a)$ and $i(t,a)$ denote the density of exposed (not infectious) and infectious individuals, respectively, at time $t\geq 0$ and demographic age $a \in [0,a^\dagger]$, and let $s^*(a)$ denote the density of susceptible individuals at equilibrium, which is determined by the vaccination rate $\nu$ (see \Cref{app:rubella} 
 for the details). 
The linearization around the disease-free equilibrium reads as follows:
\begin{equation} \label{rubella_lin}
\begin{cases}
\partial_t e(t, a)+\partial_a e(t,a)=\displaystyle\int_0^{a^\dagger} \beta(a,\xi) i(t, \xi)\dd \xi -\sigma e(t, a),\\[3mm]
\partial_t i(t, a)+\partial_a i(t,a)=\sigma e(t, a)-\gamma i(t, a), \\[3mm]
e(t, 0)=0,\\[0.5mm]
i(t, 0)=\displaystyle\int_0^{a^\dagger} b(a) i(t, a)\dd a, 
\end{cases}
\end{equation}
where $\beta(a,\xi)= s^*(a) \Pi(\xi) k(a, \xi) $ and $b(a) = qf(a)\Pi(a)$, see \Cref{tab:rubella} for more details. 

\begin{figure}
    \centering
    \includegraphics[width=1.\textwidth]{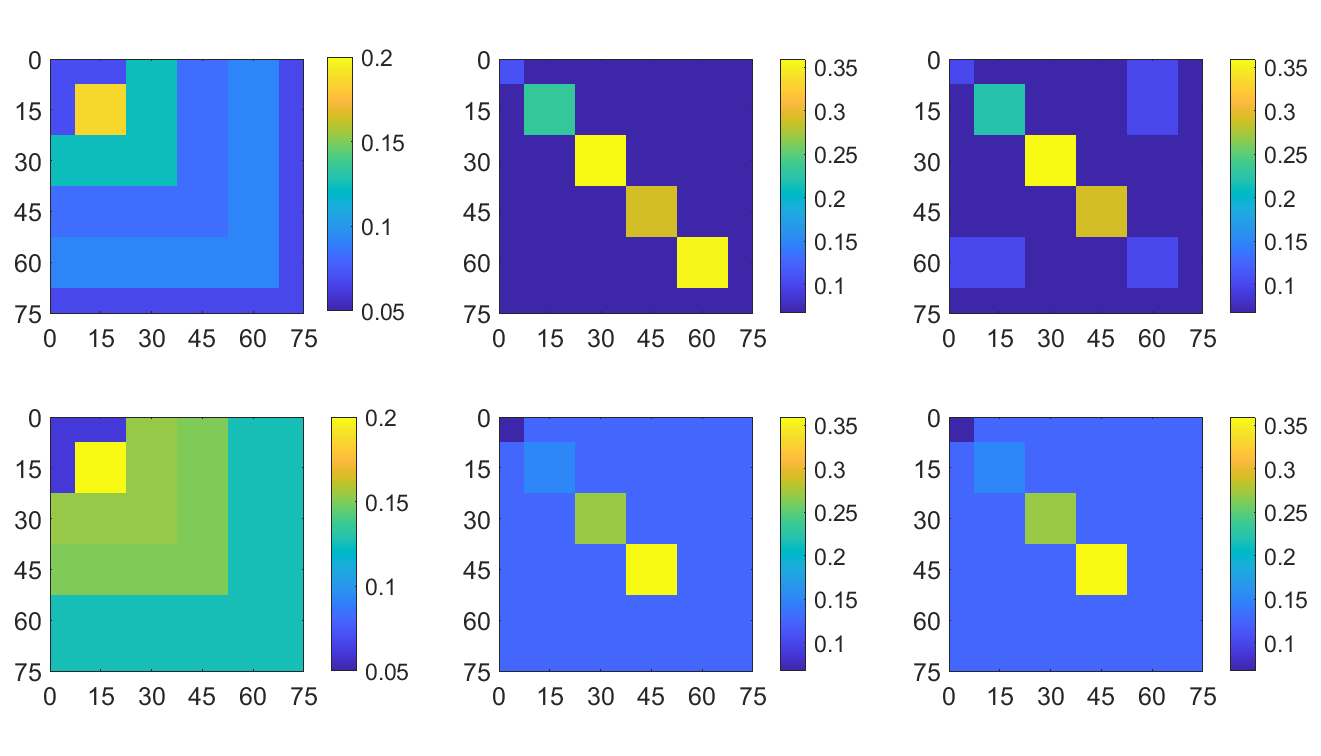}
    \caption{Model \eqref{rubella}. Estimated WAIFW matrices ($k_{ij}$) for case a (upper row) and case b (lower row). Numerical values reported in \Cref{tab:rubellaKi}.}
    \label{fig:rubellaWAIFW}
    
    \bigskip
    
    \bigskip

    \bigskip
    \includegraphics[width=.85\textwidth]{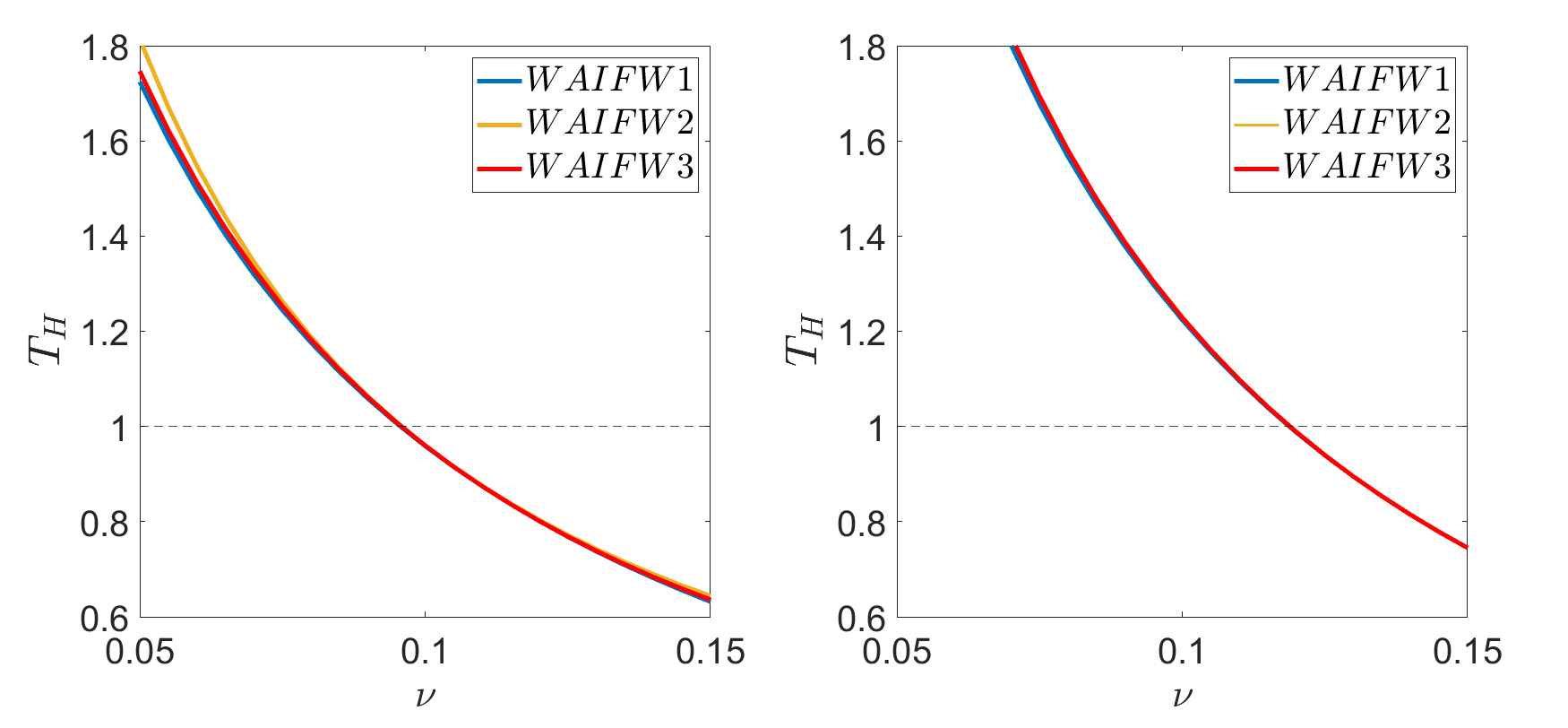}
    \caption{Model \eqref{rubella}. $T_H$ as a function of $v$ for different choices of the WAIFW-matrix, case a (left) and case b (right) according to \Cref{foirubella}.}
    \label{fig:rubellavax}
\end{figure}

Following \cite{rubella1985}, we assume that the transmission rate $k$ is piecewise constant among six age groups, i.e., 
given $0=\bar a_0<\bar a_1<\dots \bar a_6=a^\dagger$, 
\begin{equation}\label{PCW_krubella}
k(a, \xi)\equiv k_{ij}\quad \text{for}\quad(a, \xi)\in [\bar a_{i-1}, \bar a_i)\times [\bar a_{j-1}, \bar a_j),\quad  i,j=1,\dots, 6,
\end{equation}
so that we can estimate it from existing data using the well-known procedure of \cite[Appendix A]{Anderson1985} (that we recall in \Cref{app:rubella} for the reader's convenience).
In this regard, we consider force of infection data from two different datasets: one for the South East of England in 1980 (case a) and one for Leeds in 1978 (case b), as summarized in \Cref{foirubella}. These datasets fix the age class division at 5, 10, 15, 20, and 30 years of age.
The piecewise form in \eqref{PCW_krubella} gives us a \emph{Who Acquires Infection From Whom} (WAIFW) matrix $(k_{ij})_{i,j=1,\dots 6}$, which collects the contact rates between different age groups. 
We consider three different forms of the WAIFW, which capture different features in the transmission patterns. The estimated values of $k_{ij}$ are illustrated in \Cref{fig:rubellaWAIFW}. More details on the parameters and data used for the estimation are available in \Cref{app:rubella}.
 
\begin{table}[th]
\centering\small
\smallskip
\begin{tabular}{c c c c c c c}
\toprule
 & \multicolumn{2}{c}{\textbf{WAIFW1} (case a)} & \multicolumn{2}{c}{\textbf{WAIFW2} (case a)} & \multicolumn{2}{c}{\textbf{WAIFW3} (case a)} \\
\cmidrule(rl){2-3} \cmidrule(rl){4-5} \cmidrule(rl){6-7}
$v$ & {$R_0$} & {$T_H$} & {$R_0$} & {$T_H$} & {$R_0$} & {$T_H$} \\
\midrule
0  & 5.4206 & 5.4223 & 5.9228 & 5.9243 & 5.4592 & 5.4609 \\
0.05  & 1.7240 & 1.7242 & 1.8130 & 1.8133 & 1.7465 & 1.7468\\
0.09  & 1.0580 &  1.0580 & 1.0624 & 1.0625 & 1.0612 & 1.0612\\
0.10  & 0.9588 & 0.9588 & 0.9600 & 0.9600 & 0.9599 & 0.9599\\
0.15  & 0.6317 & 0.6316 & 0.6436 & 0.6435 & 0.6360 & 0.6359\\
\bottomrule
\end{tabular}
\label{tab:rubella2}
\caption{Model \eqref{rubella}. $R_0$ and $T_H$ for different values of the vaccination rate $v$ and different choices of the WAIFW matrix (case a).}

\bigskip
\begin{tabular}{c c c c c c c}
\toprule
 & \multicolumn{2}{c}{\textbf{WAIFW1} (case b)} & \multicolumn{2}{c}{\textbf{WAIFW2} (case b)} & \multicolumn{2}{c}{\textbf{WAIFW3} (case b)} \\
\cmidrule(rl){2-3} \cmidrule(rl){4-5} \cmidrule(rl){6-7}
$v$ & {$R_0$} & {$T_H$} & {$R_0$} & {$T_H$} & {$R_0$} & {$T_H$} \\
\midrule
0  & 9.5110 & 9.5141 & 9.5759 & 9.5790 & 9.5759 & 9.5790 \\
0.05  & 2.5013 & 2.5018 & 2.5376 & 2.5381 & 2.5376 & 2.5381\\
0.09  & 1.3776 & 1.3777 & 1.3866 & 1.3867 & 1.3866 & 1.3867\\
0.10  & 1.2236 & 1.2237 & 1.2289 & 1.2290 & 1.2289 & 1.2290\\
0.15  & 0.7449 & 0.7448 & 0.7446 & 0.7445 & 0.7446 & 0.7445\\
\bottomrule
\end{tabular}
\bigskip
\caption{Model \eqref{rubella}. $R_0$ and $T_H$ for different values of the vaccination rate $v$ and different choices of the WAIFW matrix (case b).}
\label{tab:rubella3}
\end{table}

\begin{figure}
    \centering
    \includegraphics[width=.8\textwidth]{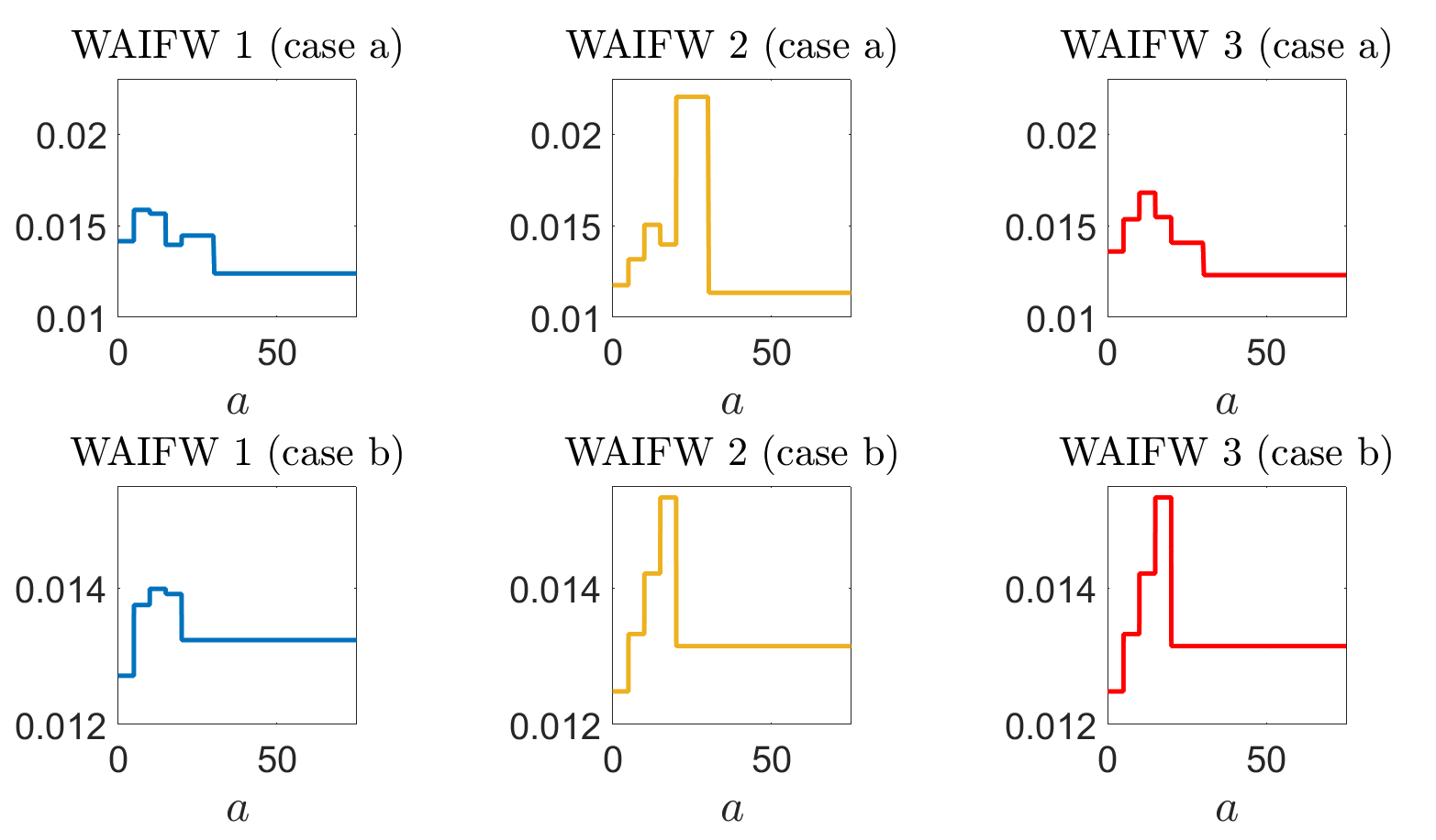}
    \caption{Model \eqref{rubella}. Approximated eigenfunctions of the type reproduction operator (see also \Cref{s3}) relevant to horizontal transmission in the absence of vaccination, for different choices of the WAIFW matrix for case a (upper row) and case b (lower row).}
    \label{fig:rubellaEIGf}

\bigskip
    \includegraphics[width=.91\textwidth]{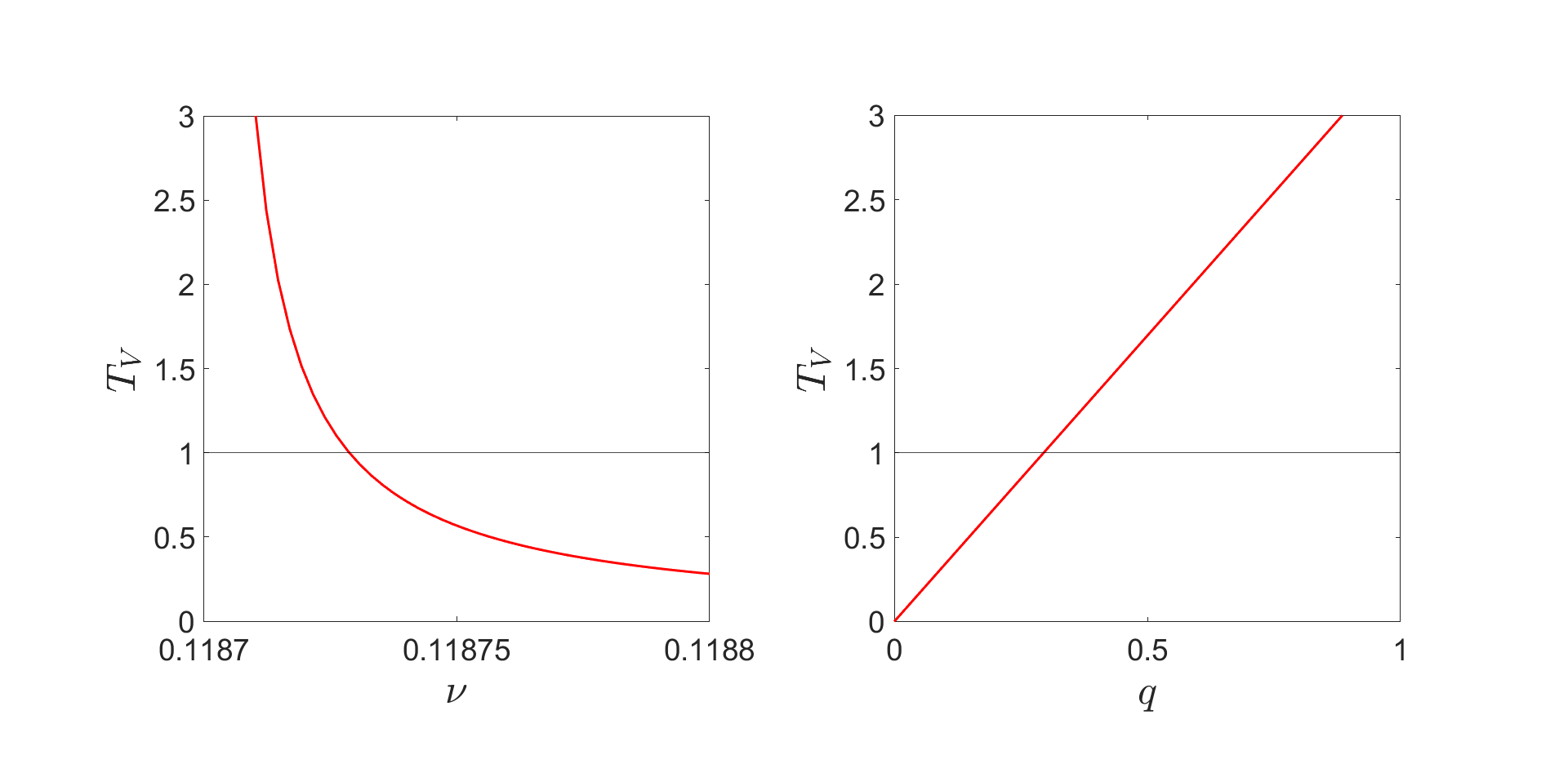}
    \caption{Model \eqref{rubella}. $T_V$ as a function of $v$ (left) and $q$ (right) for case b with WAIFW1. When not varied, the parameters are $\nu=0.11871$ and $q=0.9$.}
    \label{fig:rubellavaxtv}
\end{figure}

\bigskip
We compute the basic reproduction number, $R_0$, and the type reproduction number for horizontal transmission, $T_H$, for different choices of the vaccination rate $v$ and the WAIFW matrix.
The results are collected in \Cref{tab:rubella2} (case a) and \Cref{tab:rubella3} (case b), rounded to four decimal digits.

The computed values of $R_0$ and $T_H$ are never substantially different (we can appreciate some differences only at the third decimal digit), thus suggesting that, for these data-informed parameter values, vertical transmission has a substantially smaller effect on the epidemic spread compared to horizontal transmission.
Note that, for case b, the results obtained with WAIFW2 and WAIFW3 are identical: this is actually a consequence of the particular force of infection data in \Cref{foirubella}, which is the same for the two age groups 20--29 and 30--75 years old, see also \cite[pag. 324]{rubella1985}.
Additionally, the values in \Cref{tab:rubella2} and \Cref{tab:rubella3} numerically illustrate the known relations between $T_H$ and $R_0$, namely $0<T_H<R_0<1$ and $1<R_0<T_H$\, \cite{Inaba2012}.

Both $R_0$ and $T_H$ decrease for increasing $v$ for all choices of WAIFW matrix, see \Cref{fig:rubellavax}. This is expected since a larger vaccination rate reduces the density of the susceptible population at equilibrium. On the other hand,  different choices of the WAIFW matrix may result in slightly different quantitative values of $R_0$ and $T_H$, which reflects how different assumptions on the mixing patterns between individuals of different ages can affect transmission. This difference is even more evident looking at the eigenfunction of the type reproduction operator relevant to $T_H$ (normalized in the $L^1$-norm),
see \Cref{fig:rubellaEIGf}.

Finally, we compute the type reproduction number for vertical transmission $T_V$. \Cref{fig:rubellavaxtv} shows how $T_V$ depends on $\nu$ (left) and $q$ (right) for case b with WAIFW1. $T_V$ is a decreasing function of $\nu$ and linearly increases with $q$ (the ad hoc values of $\nu$ are chosen for illustrative purposes).

\section{Discussion and conclusions}\label{end}
In this paper, we have proposed a general numerical method to approximate the reproduction numbers of a class of age-structured population models with a finite age span. 
We presented applications to epidemic models that show how the method can compute different reproduction numbers, including the basic and the type reproduction numbers. Additionally, these examples show that, even when analytical expressions for the reproduction numbers are available, their computation may still require numerical approximations. Hence, our approach may represent an efficient and general alternative.  

\medskip To our knowledge, this is the first numerical method that can approximate the many types of reproduction numbers for any splitting of the processes into birth and transition. This flexibility is made possible by working with the equivalent formulation for the age-integrated state, which has several advantages. First, it allows us to interpret processes described by the boundary condition as perturbations of an operator with trivial domain. Second, since the state is continuous, we can work with polynomial interpolation. Moreover, the additional regularity provided by the integral mapping permits us to investigate the convergence of the approximated eigenvalues without the need to look for a characteristic equation as in \cite{BredaKuniyaRipollVermiglio2020}. In fact, we can prove under mild (and biologically meaningful) assumptions that if $\mathcal M_-$ is invertible with bounded inverse, then there exists a positive integer $\bar N$ such that $\mathcal M_{N}$ is also invertible with bounded inverse for all $N\ge \bar N$, and that the eigenvalues of $\mathcal H_N$ converge to those of $\mathcal H$ with rate that depends on the regularity of the relevant eigenfunctions. In this paper, we experimentally investigated the convergence in some cases where the reproduction numbers were known, and we refer to the work in preparation \cite{DereggiScarabelVermiglio} for the theoretical details.

\medskip
Here, we focused on models with one structuring variable, namely age. However, to obtain a more realistic portrait of the dynamics of a population, one could also be interested in considering models with two (or more) structures (e.g., demographic age and infection age), see for example \cite{KangHuoRuan2021, Webb1985b} and references therein. Unfortunately, considering an additional structuring variable brings in many difficulties in the theoretical and numerical study of these models. In fact, the hyperbolic nature of the infinitesimal generator of the semigroup makes the stability analysis more involved in the presence of discontinuities that propagate along the characteristic lines; for instance, these could be generated by corner singularities in the domain of the structuring variables, which is typically a rectangle in $\R^2$. In this context, working with the integrated state permits us to work with more regular spaces and to have an explicit expression for the infinitesimal generator, without the need for additional smoothness assumptions on the model coefficients, or compatibility assumptions on the boundary conditions as in \cite{BredaDeReggiScarabelVermiglioWu2022}. A first work in this direction is \cite{ando2023}, where the integrated state framework was used to approximate the spectrum of the infinitesimal generator relevant to the semigroup of a linear, scalar model with two structures. 
Following this idea, we plan to extend the method presented in this paper to models with more than one structure. 

\medskip
A limitation of this work is the assumption of finite age span. Hence, another interesting extension of this method regards models with unbounded structuring variables, which are common in the literature when considering probability distributions with unbounded support. For handling this problem numerically, one can either truncate the domain or resort to interpolation on exponentially weighted spaces and Laguerre-type nodes. For recent applications of these techniques to delay and renewal equations, which can also be used to model structured populations \cite{CalsinaDiekmannFarkas2016}, see \cite{GyllenbergScarabelVermiglio2018, Scarabel2023}.

\appendix

\section{Modeling details} \label{app:models}
In this appendix, we collect some further modeling and computational details regarding  the models considered in \Cref{sec:examples}, including the original formulation of the models in their nonlinear form. We use capital letters ($S$, $I$, $R$\dots) to denote the age-densities (or numbers depending on the context), and small letters to denote the densities divided by either the total size or the age distribution of the host population (e.g., $s(t,a) = S(t,a)/P(t,a)$).

\subsection{An epidemic model structured by infection age}\label{app:TSI}
Model \eqref{TSI} is obtained by partitioning the population into three classes (susceptibles, infected and removed), where only the infected class is structured by the infection age, see for example \cite[Chapter 7]{Iannelli1995} and \cite[Section 5.3]{Inaba2017}. Let $S(t)$ and $R(t)$ denote the number of susceptible and removed individuals, respectively, at time $t\ge0$, and let $I(t,\tau)$ denote the density of infected individuals at time $t\ge0$ and infection age $\tau\in[0, \tau^\dagger]$. We assume that the total population $P:=S(t)+R(t)+\int_0^{\tau^\dagger}I(t, \tau)\dd \tau$ is constant (no demographic turnover). The model reads as follows:
\begin{equation*}\label{TsiAppendix}
\begin{cases}
S'(t)=- \cfrac{S(t)}{P} \displaystyle\int_0^{\tau^\dagger} b(\tau)I(t, \tau)\dd \tau ,\\[3mm]
\partial_t I(t, \tau)+\partial_\tau I(t, \tau) =-\gamma(\tau) I(t, \tau),\\[1mm]
I(t, 0)= \cfrac{S(t)}{P} \displaystyle\int_0^{\tau^\dagger} b(\tau)I(t, \tau)\dd \tau, \\
R'(t)=\displaystyle\int_0^{\tau^\dagger}\gamma(\tau)I(t, \tau)\dd\tau,
\end{cases}
\end{equation*}
and can be rewritten in terms of the new variables
\begin{equation*}
s(t):=\cfrac{S(t)}{P},\quad i(t, \tau):=\cfrac{I(t, \tau)}{P},\quad  r(t):=\cfrac{R(t)}{P},
\end{equation*}
as
\begin{equation*}
\begin{cases}
s'(t)=- s(t) \displaystyle\int_0^{\tau^\dagger} b(\tau)i(t, \tau)\dd \tau ,\\[3mm]
\partial_t i(t, \tau)+\partial_\tau i(t, \tau) =-\gamma(\tau) i(t, \tau),\\[0.5mm]
i(t, 0)= s(t) \displaystyle\int_0^{\tau^\dagger} b(\tau)i(t, \tau)\dd \tau, \\
r'(t)=\displaystyle\int_0^{\tau^\dagger}\gamma(\tau)i(t, \tau)\dd\tau.
\end{cases}
\end{equation*}

\subsection{A multi-strain epidemic model with host age structure} \label{app:multistrain}
We consider a simplified version of the model proposed in \cite{Qiu2012}, with two classes of infected individuals structured by demographic age, describing the dynamics of two competing strains in a host population. 
Let $S(t,a)$ denote the density of susceptible individuals at time $t\ge0$ and demographic age $a\in [0, a^\dagger]$, and let $I_1(t,a)$ and $I_2(t,a)$ denote the density of infectious individuals with strain $1$ and $2$, respectively, at time $t\ge0$ and demographic age $a\in [0, a^\dagger]$. We neglect additional disease-induced mortality. The full nonlinear model reads as follows:
\begin{equation*}
\begin{cases}
\partial_t S(t, a)+\partial_a S(t, a)=-(\lambda_1(t, a)+\lambda_2(t, a)+\mu(a)) S(t, a)
,\\[3mm]
\partial_t I_1(t, a)+\partial_a I_1(t, a)=\lambda_1(t,a) S(t, a)-\mu(a) I_1(t, a)
,\\[3mm]
\partial_t I_2(t, a)+\partial_a I_2(t, a)=\lambda_2(t,a) S(t, a)-\mu(a) I_2(t, a)
,\\[0.5mm]
S(t, 0) = R_0^d \Phi(Q(t)) \displaystyle\int_0^{a^\dagger} f(a) P(t,a) \dd a
,\\[3mm]
I_1(t, 0) = I_2(t, 0) = 0,
\end{cases}
\end{equation*}
where $Q(t) := \int_0^{a^\dagger} P(t,\xi)\dd \xi$,  $P(t,a) := S(t,a)+I_1(t,a)+I_2(t,a)$ is the age distribution of the host population, and the force of infection $\lambda_j$ satisfies
$$ \lambda_j(t,a) = K_j(a) \int_0^{a^\dagger} q_j(\xi) I_j(t , \xi) \dd \xi, \quad j=1,\,2,$$
where $q_j$ is the age-specific infection rate for strain $j$, and $K_j$ is the age-specific susceptibility of susceptible individuals to strain $j$.
For the demographic process, $P(t,a)$ is assumed to be at demographic equilibrium, i.e., the death rate $\mu$ and the per capita fertility rate $f$ are such that $\int_0^{a^\dagger} f(a) \Pi(a) \dd a = 1$, where 
\begin{equation} \label{survival}
\Pi(a) := \mathrm{e}^{-\int_0^a \mu(\xi) \dd \xi}
\end{equation}
is the survival probability. If $R_0^d>1$, then there exists an endemic equilibrium such that the population profile at equilibrium, $P^*(a)$, satisfies
\begin{equation}\label{Pstar}
P^*(a) = \frac{\Pi(a) P_0}{\int_0^{a^\dagger} \Pi(\xi) \dd \xi},
\end{equation}
for $P_0 = \int_0^{a^\dagger} P^*(a) \dd a = \Phi^{-1}(1/R_0^d)$.

To simplify the model, we define the variables
$$ s(t,a) := \frac{S(t,a)}{P^*(a)}, \quad i_j(t,a) := \frac{I_j(t,a)}{P^*(a)}, \quad j=1,\,2, $$
and obtain the following nonlinear system of equations:
\begin{equation*}
\begin{cases}
\partial_t s(t, a)+\partial_a s(t, a)= -(\lambda_1(t, a)+\lambda_2(t, a)) s(t, a)
,\\[3mm]
\partial_t i_1(t, a)+\partial_a i_1(t, a)= \lambda_1(t,a) s(t, a)
,\\[3mm]
\partial_t i_2(t, a)+\partial_a i_2(t, a)= \lambda_2(t,a) s(t,a)
,\\[3mm]
s(t,0) = 1
,\\[3mm]
i_1(t, 0) = i_2(t, 0) = 0,
\end{cases}
\end{equation*}
where the force of infection can be written (equivalently) as follows:
$$ \lambda_j(t,a) = K_j(a) \int_0^{a^\dagger} q_j(\xi) P^*(\xi) i_j(t, \xi) \dd \xi, \quad j=1,\,2.$$

The linearization around the disease-free equilibrium is reported in \eqref{multistrain}. Regarding the boundary equilibria $E_1^*=(s^*_1,i^*_1,0)$ and $E_2^*=(s^*_2,0,i^*_2)$, where only one strain is present in the population, they satisfy, for $k=1,2$,
\begin{equation} \label{end_equil}
\frac{\dd s^*_k(a)}{\dd a} = -\lambda_k^*(a) s^*_k(a),  \qquad  \frac{\dd i^*_k(a)}{\dd a} = \lambda^*_k(a) s^*_k(a), \quad a \in [0,a^\dagger],
\end{equation}
with $s^*_k(0)=1$, $i^*_k(0)=0$, and 
$\lambda^*_j(a) = K_j(a) \int_0^{a^\dagger} q_j(\xi) P^*(\xi) i^*_j(\xi) \dd \xi$, $j=1,\,2.$

\paragraph{Numerical solution of the endemic equilibrium.} Note that system \eqref{end_equil} for $s_k^*$ and $i_k^*$ cannot be solved analytically. To compute the equilibrium values to perform the numerical tests in the main text, the solutions were approximated numerically. Consider the equilibrium $E_1^*$ for illustrative purposes. For $n\in \N$, we discretize the interval $[0,a^\dagger]$ using Chebyshev extremal nodes $\{0=a_0< a_1< \dots< a_n=a^\dagger \}$. Then, the derivative at each node is approximated using spectral differentiation, and the integral is approximated using Clenshaw--Curtis quadrature formulas with weights $w_{n,j}$.
Let $S_n,I_n \in \R^{n+1}$ be two vectors such that, for $j=0,\dots,n$, each entry $S_{n,j}$ approximates $s^*_1(a_j)$, and each entry $I_{n,j}$ approximates $i^*_1(a_j)$. Let $D \in \mathbb{R}^{(n+1)\times (n+1)}$ be the differentiation matrix associated with the nodes. Finally, let $K_n, Q_n \in \R^{n+1}$ such that $K_{n,j} = K_1(a_j)$, and $Q_{n,j} = w_{n,j} \, q_1(a_j) P^*(a_j)$. Then, we can write the following approximating system for the unknowns $S_n,I_n$:
\begin{equation*}\begin{cases}
D S_n = - (Q_n \cdot I_n ) K_n * S_n, \\[2mm]
D I_n = (Q_n \cdot I_n ) K_n * S_n,
\end{cases}
\end{equation*}
where $*$ denotes the element-wise product. In practice, to facilitate the convergence to the nontrivial solution, we divide both terms by $(Q_n \cdot I_n)$ and solve the corresponding system.

\subsection{A model for the spread of Rubella with vertical transmission} \label{app:rubella}
We consider a model inspired by \cite{rubella1985}. 
Let $M(t,a), S(t, a), E(t,a), I(t,a)$ and $Z(t,a)$ denote the density of individuals who are protected by maternal antibodies, susceptible, infected but not infectious, infectious, and immune (acquired naturally or via vaccination), respectively, at time $t\ge 0$ and demographic age $a\in [0, a^\dagger]$. The model reads as follows:
\begin{equation*}
\begin{cases}
\partial_t M(t, a)+\partial_a M(t,a)=-(\mu(a)+\eta)M(t, a),\\[3mm]
\partial_t S(t, a)+\partial_a S(t,a)=\eta M(t, a)-(\mu(a)+\lambda(t,a)+v(a))S(t, a),\\[3mm]
\partial_t E(t, a)+\partial_a E(t,a)=\lambda(t, a)S(t, a)-(\mu(a)+\sigma)E(t, a),\\[3mm]
\partial_t I(t, a)+\partial_a I(t,a)=\sigma E(t, a)-(\mu(a)+\gamma)I(t, a),\\[3mm]
\partial_t Z(t, a)+\partial_a Z(t,a)=\gamma I(t, a)+v(a)S(t, a)-\mu(a) Z(t,a),
\end{cases}
\end{equation*}
with the following boundary conditions:
\begin{align*}
M(t, 0)&=\displaystyle\int_0^{a^\dagger}f(a)(M(t,a)+Z(t,a))\dd a,\\
S(t,0)&=\displaystyle\int_0^{a^\dagger}f(a)[S(t,a)+E(t,a)+(1-q)I(t,a)]\dd a,\\
I(t, 0)&=q\displaystyle\int_0^{a^\dagger}f(a)I(t,a)\dd a,\\[3mm]
E(t, 0)&=Z(t, 0)=0,
\end{align*}
where
\begin{equation*}
\lambda(t, a):=\displaystyle\int_0^{a^\dagger} \hat k(a, \xi)I(t,\xi)\dd\xi
\end{equation*}
is the force of infection, for $\hat k(a, \xi)$ the transmission rate between one individual of age $a$ and one individual of age $\xi$. 
We refer to \Cref{tab:rubella} for the interpretation of the parameters.
Note that an individual is assumed to have permanent immunity once infected.

Following \cite[Chapter 6]{Inaba2017}, we assume that $\int_{0}^{a^\dagger}f(a)\Pi(a)\dd a=1$,
where $\Pi$ is the survival probability defined in \eqref{survival},
and that the age density of the host population $P(t, a):=M(t,a)+S(t,a)+E(t,a)+I(t,a)+Z(t,a)$ has already attained the stable age distribution $P(t,a)=P^*(a)$ defined in \eqref{Pstar}, for some $P_0>0$, see for example \cite[Chapter 6]{Inaba2017}.
For convenience, we define the standardized transmission rate as follows:
\begin{equation*}
    k(a,\xi) := \frac{P_0 \hat{k}(a,\xi)}{\int_0^{a^\dagger} \Pi(\theta) \dd \theta}. 
\end{equation*}
Then, if we consider new variables
\begin{gather*}
m(t, a):=\cfrac{M(t, a)}{P^*(a)},\quad  s(t, a):=\cfrac{S(t, a)}{P^*(a)},\quad e(t, a):=\cfrac{E(t, a)}{P^*(a)},\quad
i(t, a):=\cfrac{I(t, a)}{P^*(a)},\quad  z(t, a):=\cfrac{Z(t, a)}{P^*(a)},
\end{gather*}
we can reduce to the following model:
\begin{equation*}
\begin{cases}
\partial_t m(t, a)+\partial_a m(t,a)=-\eta \, m(t, a),\\[3mm]
\partial_t s(t, a)+\partial_a s(t,a)=\eta \, m(t, a)-(\lambda(t,a)+v(a))s(t, a),\\[3mm]
\partial_t e(t, a)+\partial_a e(t,a)=\lambda(t, a)s(t, a)-\sigma e(t, a),\\[3mm]
\partial_t i(t, a)+\partial_a i(t,a)=\sigma e(t, a)-\gamma i(t, a),\\[3mm]
\partial_t z(t, a)+\partial_a z(t,a)=\gamma i(t, a)+v(a)s(t, a),\\
\end{cases}
\end{equation*}
with the following boundary conditions:
\begin{align*}
m(t, 0)&=\displaystyle\int_0^{a^\dagger}f(a)\Pi(a)(m(t,a)+z(t,a))\dd a,\\
s(t,0)&=\displaystyle\int_0^{a^\dagger}f(a)\Pi(a)\left[s(t,a)+e(t,a)+(1-q)i(t,a)\right]\dd a,\\
i(t, 0)&=q\displaystyle\int_0^{a^\dagger}f(a)\Pi(a)i(t,a)\dd a,\\[3mm]
e(t, 0)&=z(t, 0)=0,
\end{align*}
where the force of infection can be expressed as follows:
\begin{equation*}
\lambda(t,a)=\int_0^{a^\dagger}k(a, \xi) \Pi(a) i(t, \xi)\dd \xi.
\end{equation*}
The disease-free equilibrium $E^*:=(m^*(a), s^*(a), e^*(a), i^*(a), z^*(a))$ explicitly reads as follows:
\begin{align*}
m^*(a)&=\left(1-s^*(0)\right) {\e}^{-\eta a},\\[2mm]
s^*(a)&=s^*(0)\,{\e}^{-\int_0^a v(\xi)\dd\xi}+\eta\int_0^a{\e}^{-\int_\xi^a v(\theta)\dd\theta}m^*(\xi)\dd \xi
,\\
 z^*(a)&=\int_0^a v(\xi)s^*(\xi)\dd \xi,\\[3mm]
e^*(a)&=i^*(a)=0,
\end{align*}
where 
\begin{equation*}
s^*(0)=\frac{\eta\int_0^{a^\dagger}f(a)\Pi(a)\int_0^a{\e}^{-\int_\xi^a v(\theta)\dd\theta}{\e}^{-\eta	\xi}\dd \xi \dd a}{1-\int_0^{a^\dagger}f(a)\Pi(a){\e}^{-\int_0^av(\theta)\dd \theta}\dd a+\eta\int_0^{a^\dagger}f(a)\Pi(a)\int_0^a{\e}^{-\int_\xi^a v(\theta)\dd\theta}{\e}^{-\eta\xi}\dd \xi\dd a}.
\end{equation*}
\smallskip
Assuming a constant vaccination rate $v(a)\equiv v$ and $\eta>v$, the density of susceptible population reads as follows:
$$s^*(a)=s^*(0)\left[{\e}^{-va}-\frac{\eta}{v-\eta}({\e}^{-\eta a}-{\e}^{-v a})\right]+\frac{\eta}{v-\eta}({\e}^{-\eta a}-{\e}^{-va}),$$
where 
$$s^*(0)=\frac{\eta(v-\eta)^{-1}\int_0^{a^\dagger}f(a)\Pi(a)({\e}^{-\eta a}-{\e}^{-v a})\dd a}{1-\int_0^{a^\dagger}f(a)\Pi(a)\left({\e}^{-va}-\eta(v-\eta)^{-1}({\e}^{-\eta a}-{\e}^{-v a})\right)\dd a}.$$ Observe that we have $s^*(a)\equiv 1$  for $v\equiv 0$.
The linearization around the disease-free equilibrium is given in \eqref{rubella}. 

\begin{table}[tp]
\begin{center}\small
\begin{tabular}{c | c  c  c c c c}
\rowcolor{gray!20}
Age class (years) & $0-4$  & $5-9$ & $10-14$ & $15-19$ & $20-29$ & $30-75$ \\
\hline
$\lambda_i$ (case a)& $0.081$ &  $0.115$ & $0.115$ & $0.083$ & $0.091$ & $0.067$ \\
$\lambda_i$ (case b)& $0.089$ &  $0.134$ & $0.151$ & $0.148$ & $0.126$ & $0.126$ \\
\hline
\end{tabular}
\label{foirubella}
\caption{Age-specific forces of infection $\lambda_i$'s (yr$^{-1}$) for \eqref{rubella_lin}. Data are taken from \cite[Table 2]{rubella1985} according to their comments at page 324.}
\end{center}
\end{table}

\smallskip
We estimate $k$ using real-world prevalence data taken from \cite[Table 2]{rubella1985} and their comments at page 324, which are collected in \Cref{foirubella}. To do this, we assume that $k$ is piecewise constant in the six age groups defined in \Cref{foirubella}, i.e.,
$$k(a, \xi)\equiv k_{ij}\quad \text{for}\quad(a, \hat a)\in [\bar a_{i-1}, \bar a_i)\times [\bar a_{j-1}, \bar a_j),\quad  i,j=1,\dots, 6.$$
%
We assume three different structures for the WAIFW matrix $(k_{ij})_{i,j=1,\dots 6}$, which are listed below:

\medskip
\begin{center}
\footnotesize
\begin{tabular}{c | c c c c c c}
\multicolumn{7}{c}{\normalsize{WAIFW1}} \\[5pt]
\specialcell{age\\class} & 1 & 2 & 3 & 4 & 5 & 6 \\
\hline
 1 & \textcolor{amber}{$k_1$}  & \textcolor{amber}{$k_1$} & \textcolor{red}{$k_3$} & \textcolor{aogreen}{$k_4$} & \textcolor{blue}{$k_5$} & \textcolor{violet}{$k_6$} \\
 2 & \textcolor{amber}{$k_1$} &  \textcolor{orange}{$k_2$} & \textcolor{red}{$k_3$} & \textcolor{aogreen}{$k_4$} & \textcolor{blue}{$k_5$} & \textcolor{violet}{$k_6$} \\
 3 & \textcolor{red}{$k_3$} &  \textcolor{red}{$k_3$} & \textcolor{red}{$k_3$} & \textcolor{aogreen}{$k_4$} & \textcolor{blue}{$k_5$} & \textcolor{violet}{$k_6$} \\
 4 & \textcolor{aogreen}{$k_4$} &  \textcolor{aogreen}{$k_4$} & \textcolor{aogreen}{$k_4$} & \textcolor{aogreen}{$k_4$} & \textcolor{blue}{$k_5$} & \textcolor{violet}{$k_6$} \\
 5 & \textcolor{blue}{$k_5$} &  \textcolor{blue}{$k_5$} & \textcolor{blue}{$k_5$} & \textcolor{blue}{$k_5$} & \textcolor{blue}{$k_5$} & \textcolor{violet}{$k_6$} \\
 6 & \textcolor{violet}{$k_6$} &  \textcolor{violet}{$k_6$} & \textcolor{violet}{$k_6$} & \textcolor{violet}{$k_6$} & \textcolor{violet}{$k_6$} & \textcolor{violet}{$k_6$} \\
\end{tabular}\hfill
\\
\bigskip
\begin{tabular}{c | c c c c c c}
\multicolumn{7}{c}{\normalsize{WAIFW2}} \\[5pt]
\specialcell{age\\class} & 1 & 2 & 3 & 4 & 5 & 6 \\
\hline
 1 & \textcolor{amber}{$k_1$}  & \textcolor{violet}{$k_6$} & \textcolor{violet}{$k_6$} & \textcolor{violet}{$k_6$} & \textcolor{violet}{$k_6$} & \textcolor{violet}{$k_6$} \\
 2 & \textcolor{violet}{$k_6$} &  \textcolor{orange}{$k_2$} & \textcolor{violet}{$k_6$} & \textcolor{violet}{$k_6$} & \textcolor{violet}{$k_6$} & \textcolor{violet}{$k_6$} \\
 3 & \textcolor{violet}{$k_6$} &  \textcolor{violet}{$k_6$} & \textcolor{red}{$k_3$} & \textcolor{violet}{$k_6$} & \textcolor{violet}{$k_6$} & \textcolor{violet}{$k_6$} \\
 4 & \textcolor{violet}{$k_6$} &  \textcolor{violet}{$k_6$} & \textcolor{violet}{$k_6$} & \textcolor{aogreen}{$k_4$} & \textcolor{violet}{$k_6$} & \textcolor{violet}{$k_6$} \\
 5 & \textcolor{violet}{$k_6$} &  \textcolor{violet}{$k_6$} & \textcolor{violet}{$k_6$} & \textcolor{violet}{$k_6$} & \textcolor{blue}{$k_5$} & \textcolor{violet}{$k_6$} \\
 6 & \textcolor{violet}{$k_6$} &  \textcolor{violet}{$k_6$} & \textcolor{violet}{$k_6$} & \textcolor{violet}{$k_6$} & \textcolor{violet}{$k_6$} & \textcolor{violet}{$k_6$} \\
\end{tabular}\hfill
\\
\bigskip
\begin{tabular}{c|cccccc}
\multicolumn{7}{c}{\normalsize{WAIFW3}} \\[5pt]
\specialcell{age\\class} & 1 & 2 & 3 & 4 & 5 & 6 \\
\hline
 1 & \textcolor{amber}{$k_1$}  & \textcolor{violet}{$k_6$} & \textcolor{violet}{$k_6$} & \textcolor{violet}{$k_6$} & \textcolor{blue}{$k_5$} & \textcolor{violet}{$k_6$} \\
 2 & \textcolor{violet}{$k_6$} &  \textcolor{orange}{$k_2$} & \textcolor{violet}{$k_6$} & \textcolor{violet}{$k_6$} & \textcolor{blue}{$k_5$} & \textcolor{violet}{$k_6$} \\
 3 & \textcolor{violet}{$k_6$} &  \textcolor{violet}{$k_6$} & \textcolor{red}{$k_3$} & \textcolor{violet}{$k_6$} & \textcolor{violet}{$k_6$} & \textcolor{violet}{$k_6$} \\
 4 & \textcolor{violet}{$k_6$} &  \textcolor{violet}{$k_6$} & \textcolor{violet}{$k_6$} & \textcolor{aogreen}{$k_4$} & \textcolor{violet}{$k_6$} & \textcolor{violet}{$k_6$} \\
 5 & \textcolor{blue}{$k_5$} &  \textcolor{blue}{$k_5$} & \textcolor{violet}{$k_6$} & \textcolor{violet}{$k_6$} & \textcolor{blue}{$k_5$} & \textcolor{violet}{$k_6$} \\
 6 & \textcolor{violet}{$k_6$} &  \textcolor{violet}{$k_6$} & \textcolor{violet}{$k_6$} & \textcolor{violet}{$k_6$} & \textcolor{violet}{$k_6$} & \textcolor{violet}{$k_6$} \\
\end{tabular}
\end{center}

\medskip
In \Cref{tab:rubellaKi}, we list the values of $k_i$, $i=1,\dots,6$, obtained from \Cref{foirubella} using the procedure described below. 

\begin{table}[h]
\centering\small
\begin{tabular}{c c c c c c c}
\toprule
 & \multicolumn{2}{c}{\textbf{WAIFW1}} & \multicolumn{2}{c}{\textbf{WAIFW2}} & \multicolumn{2}{c}{\textbf{WAIFW3}} \\
\cmidrule(rl){2-3} \cmidrule(rl){4-5} \cmidrule(rl){6-7}
 & (case a) & (case b) & (case a) & (case b) & (case a) & (case b) \\
\midrule
$k_1$  & 2.205 & 1.938 & 3.467 & 0.730 & 3.198 & 0.730 \\
$k_2$ & 5.905 & 6.502 & 7.356 & 4.811 & 7.049 & 4.811\\
$k_3$  & 3.932 & 4.866 & 11.415 & 8.565 & 11.415 & 8.565\\
$k_4$ & 2.644 & 4.753 & 9.212 & 12.664 & 9.212 & 12.664\\
$k_5$ & 2.940 & 4.000 & 11.275 & 4.000 & 3.208 & 4.000\\
$k_6$ & 2.132 & 4.000 & 2.132 & 4.000 & 2.132 & 4.000\\
\bottomrule
\end{tabular}
\label{tab:rubellaKi}
\caption{Values of $k_i$, $i=1,\dots, 6$ in case a and  case b, estimated from the force of infection data in \Cref{foirubella} with different configurations of the WAIFW matrix.}
\end{table}

\paragraph{Estimation of age-dependent transmission rates.}\label{estimatek}
Here, we recall the procedure described in \cite[Appendix A]{Anderson1985} to estimate the age-dependent transmission rate $k$ under the hypothesis that it is piecewise constant among different age groups, i.e., 
$$k(a, \xi)\equiv k_{ij}\quad \text{for}\quad(a, \xi)\in [\bar a_{i-1}, \bar a_i]\times [\bar a_{j-1}, \bar a_j],\quad  i,j=1,\dots, n,$$
for given $0=\bar a_0<\bar a_1<\dots <\bar a_n=a^\dagger$. 
We assume that the age-specific mortality rate $\mu$ has the following form:
\begin{equation*}
\mu(a)=\begin{cases} 0,\quad&\text{if }a\le a^\dagger,\\ \infty &\text{otherwise},
\end{cases}
\end{equation*}
so that $\int_0^{a^\dagger} \Pi(a) \dd a = a^\dagger $, and that the age-specific force of infection
\begin{equation}\label{PCWFOIrubella}
\lambda(a)\equiv \lambda_i,\quad\text{for}\quad a\in [\bar a_{i-1}, \bar a_i],\ i=1,\dots, n,
\end{equation}
is known.
 
Then, the following algorithm can be applied:
\begin{enumerate}
    \item define $\psi_0:= 0$ and $\psi_i:=\sum_{j=1}^i \lambda_j(\bar a_j-\bar a_{j-1})$ for $i=1,\dots, n$;
    \item define $\Psi_i:=\exp\left(-\psi_{i-1}\right)-\exp\left(-\psi_{i}\right)$  for $i=1,\dots, n$;
    \item\label{3} solve the linear problem $\lambda_i= \frac{1}{\gamma} \sum_{j=1}^n k_{ij}\Psi_j$  for $i=1,\dots, n$.
\end{enumerate}
Observe that the linear system in \ref{3} is over-determined; thus some hypotheses on the structure of the WAIFW matrix $(k_{ij})_{i,j=1,\dots, n}$ are needed. We refer the reader to \cite{Anderson1985} for some possible choices.

\section*{Acknowledgments}
We thank the reviewers for their valuable comments, which helped us to improve the presentation of the paper.

The authors are members of the INdAM Research group GNCS and of the UMI Research group ``Mo\-del\-li\-sti\-ca so\-cio-epi\-de\-mio\-lo\-gi\-ca''.
The work of Simone De Reggi and Rossana Vermiglio was supported by the Italian Ministry of University and Research (MUR) through the PRIN 2020 project (No.\ 2020JLWP23) ``Integrated Mathematical Approaches to Socio-E\-pi\-de\-mio\-lo\-gi\-cal Dynamics'', Unit of Udine (CUP: G25F22000430006). Francesca Scarabel acknowledges a Heilbronn Small Grant Call award by HIMR that partly funded a research visit during which some of this research was conducted. 

\section*{Use of AI tools declaration}
The authors declare they have not used Artificial Intelligence (AI) tools in the creation of this article.

\section*{Code availability}
MATLAB demos are available at GitLab via \url{https://cdlab.uniud.it/software}.

 \printbibliography

\end{document}